\documentclass[12pt]{iopart}

\newcommand{\beq}{\begin{equation}}
\newcommand{\eneq}{\end{equation}}
\newcommand{\be}{\begin{equation}}
\newcommand{\ee}{\end{equation}}
\newcommand{\bea}{\begin{eqnarray}}
\newcommand{\eea}{\end{eqnarray}}

\usepackage{epsfig}
\usepackage{graphicx}
\begin{document}

\title{Josephson current through a long quantum wire}

\author{Domenico Giuliano$^{1}$ and Ian Affleck$^{2}$}

\address{$^{1}$
Dipartimento di Fisica, Universit\`a della Calabria Arcavacata di Rende I-87036, Cosenza, Italy
and
I.N.F.N., Gruppo collegato di Cosenza, Arcavacata di Rende I-87036, Cosenza, Italy\\
$^{2}$ Department of Physics and Astronomy, University of British 
Columbia, Vancouver, B.C., Canada, V6T 1Z1}
\ead{ \\
$^{1}$ domenico.giuliano@fis.unical.it \\
$^{2}$ iaffleck@phas.ubc.edu}
\begin{abstract}
The dc Josephson current through a long SNS junction receives contributions from 
both Andreev bound states localized in the normal region as well as from scattering 
states incoming from the superconducting leads. We show that in the limit 
of a long junction, this current, at low temperatures, can be expressed entirely in terms of 
properties of the Andreev bound states at the Fermi energy: the normal and Andreev 
reflection amplitudes at the left-hand and at the right-hand S-N interface. This has important implications for 
treating interactions in such systems.

\end{abstract}

\maketitle

\section{Introduction}
\label{intro}

As was shown by Josephson \cite{Andreec} a current can pass between two superconductors separated by a normal material, 
even with zero potential difference. At temperature $T$, 
this Josephson current is determined by the difference of the phase of 
the order parameter in the two superconductors, $\chi$:
\be I [ \chi ; T  ] =2e{dF\over d\chi}\ee
where $F$ is the free energy \cite{bodeg}.  Using the Bardeen-Cooper-Schrieffer (BCS) approximation in the superconducting leads and 
ignoring interactions in the normal region, $I(\chi )$ can be 
expressed as a sum over single quasi-particle energy levels, $E_n$:
\be I [ \chi ; T ] =2e\sum_nf(E_n){dE_n\over d\chi}\label{I}\ee
where $f(E)=1/[e^{E/T}+1]$ is the Fermi function and we measure energies from the chemical potential. 
In general, these states are of at least two distinct forms.  There are Andreev bound 
states (ABS) \cite{Andreev}, with energies $|E|<\Delta$, where $\Delta$ is the superconducting gap,
 which are localized in the normal region and whose wavefunctions
decay exponentially into the superconducting leads. 
There are also scattering states (SS), with energies $|E|>\Delta$ corresponding to 
waves coming in from infinity in the superconducting leads and being reflected 
and transmitted. If the bottom of the band in the normal material is lower than 
the bottom of the band in the superconducting leads, there are, in addition, normal bound states, localized in 
the normal region which also decay exponentially in the leads. 
As remarked in  \cite{kulik}  and later on discussed in detail in  \cite{ishi}  (where,
in the limit in which normal reflection processes at the S-N interfaces can be
neglected,   SS's are correctly taken into account, fixing the result of  \cite{kulik} ),
in general, all  types of states contribute to the Josephson current. 
In addition, it is worth stressing that, summing over all the states, to 
get Eq. (\ref{I}), may be quite a difficult  task to achieve,  since it turns out that the net current contains  very small 
differences between very large terms \cite{bardeen}. 

A convenient way to compute Eq. (\ref{I}) is to express the total current, 
summed over all types of states, as a contour integral in the 
complex energy plane, involving the $S$-matrix.  In particular, using  
an adapted version of the formalism developed in  \cite{been.0,furus}, we
will show that,  making a minimal set
of reasonable assumptions about the properties of the $S$-matrix, on appropriately deforming
the integration path, one may write the dc Josephson current as a sum over
Matsubara frequencies which, when $T \to 0$, turns into   
an integral over the imaginary axis. This allows for getting rid of the wild oscillations
in the integrand function arising at real values of the energy, thus paving the way to
 a systematic analysis of the long junction limit. 

The limit of a long narrow normal region was considered in  \cite{ACZ} , using a 
nearest neighbor tight-binding model and initially ignoring interactions. In particular, it was 
assumed that the length of the normal region, $\ell$, was much greater than the coherence length, 
or equivalently than the finite size gap, $\pi v_F/\ell \ll \Delta$ where $v_F$ is the Fermi velocity 
in the normal region. Furthermore, only $T=0$ was considered. In this limit it appears natural 
to integrate out the gapped superconductors and derive an effective Hamiltonian for the normal 
region, with local pairing interactions induced by the proximity effect at its boundaries. 
Such an effective Hamiltonian was used to derive the Josephson current. In this approach, 
only ABS's are considered. Due to a remarkable cancellation between pairs of ABS's it was 
found that  the current, to order $1/\ell$, could be expressed in terms of scattering amplitudes
 at the Fermi energy only.  

This approach was called into  question by the results of  \cite{Perfetto}.  There it was verified that the ABS's gave 
the entire Josephson current for long junctions in the unrealistic limit $\Delta > 2 J$ where $4J$ is the bandwidth 
in the normal region. However, numerical results for intermediate length junctions seemed to suggest a significant 
contribution from SS's for  $\Delta / (2 J) < 1$.  

In this paper we study general models of long non-interacting SNS junctions without integrating out the superconducting 
leads.  We prove that, for $v_F/\ell$ and $T\ll \Delta$, the Josephson current can indeed be expressed in terms of data 
 at the Fermi level only.  We emphasize that states far from the Fermi energy make large contributions to the current; it is 
just that these nearly cancel for large $\ell$, leading to our main formula for the dc Josephson current

\beq
I^{(0)} [ \chi ] = - \frac{4 e v_F}{\pi \ell} \partial_\chi \vartheta^2 ( \chi ) 
\:\:\:\: , 
\label{main}
\eneq
\noindent
with $\vartheta ( \chi )$ being a real function of $\chi$, defined by

\beq
\vartheta ( \chi ) = {\rm arccos} \{  \hbox{Re} [  \bar{N}_R^p \bar{N}_L^p e^{ 2 i \alpha_F \ell}  
+  \bar{A}_R^p \bar{A}_L^h   ] \}
\label{main.1}
\:\:\:\: .
 \eneq
\noindent
In Eq. (\ref{main.1})  $\bar{N}_{R/L}^{p/h} , \bar{A}_{R/L}^{p/h}$ are respectively the normal and the Andreev
single-particle/hole reflection amplitudes at the right/left-hand S-N interface
{\it evaluated at the Fermi level only}, while $\alpha_F$ is the single-particle Fermi momentum
in the central region. (In the following, we will denote by $\alpha_{p/h}$ the
single-particle/hole momentum within the central region, respectively.)
In particular, we apply   our general result 
in Eq. (\ref{main}) to the ``Blonder-Tinkham-Klapwijk (BTK) model'' \cite{btk}, obtaining an explicit 
formula for the current for $v_F/\ell \ll \Delta$. 
Then we show that our approach may readily be
extended to tight-binding models, such as the one discussed in \cite{ACZ}, whose results 
for the   current we recover when $T=0$ and $v_F/\ell \ll \Delta$.
We also extend our contour methods to finite $T$, by expressing the resulting current in terms of a 
sum along the imaginary energy axis at the Matsubara frequencies, 
 $E=i\omega_n\equiv i2\pi (n+1/2)T$. As a result, we find that the current vanishes exponentially 
when $T\gg v_F/\ell$.   

This finding is important because integrating out the superconductors provides a powerful method for 
including interaction effects in the normal region, based on boundary conformal field theory techniques \cite{ACZ}. 
(See also \cite{Titov}.)  While 
\cite{ACZ,Titov} only considered the dc Josephson current, the techniques introduced there can be 
extended \cite{GA} to the ac case by allowing for the phase of the boundary pairing interactions to evolve linearly in time, 
$\chi = e W t$, where $W$ is the voltage difference. A possible experimental realization of such a long 
SNS junction might be provided by a carbon nanotube between bulk superconductors. Using  
$v_F\approx 8.1 \times 10^5$ m/s,  $\pi v_F/\ell \approx .5 $ meV for $\ell$= 3 microns. Thus, obtaining  sufficiently 
long clean nanotubes coupled to sufficiently high $T_c$ superconductors to satisfy $\pi v_F/\ell \ll \Delta$ may 
be near the limits of current nanotechnology.

The paper is organized as follows:

\begin{itemize}
\item{ In section \ref{general}, we employ a convenient version of the
$S$-matrix approach, to  derive the general formula for the dc Josephson current across an SNS junction.}
\item{In section \ref{long} we apply the general formula to the specific case of a long SNS junction. We 
recast the final result in a systematic expansion in powers of $\ell^{-1}$ and, finally, derive
Eq. (\ref{main}) for the dc Josephson current.}
\item{ In section \ref{explicit}, we use Eq. (\ref{main}) to compute the dc Josephson current 
in the continuum  BTK  model  \cite{btk}  and in the lattice tight-binding model for the
SNS junction \cite{ACZ}.}
\item{In section \ref{fin_t} we discuss the generalization of our results to a finite
temperature $T$.}
\item{Section \ref{concl} contains conclusions.}
\item{In the appendices, we provide mathematical details of our derivation.}
\end{itemize}

\section{The general formula for the dc Josephson current }
\label{general}

To derive  a general formula for the dc Josephson current across the SNS junction, we
have to carefully sum over  contributions from both ABS's, as well as SS's \cite{kulik,ishi}. 
An effective way of performing the sum over both sets of states is provided by the $S$-matrix,
approach, which we extensively discuss in the following. The $S$-matrix approach 
has been showed  to be quite useful in studying superconducting point contacts, as it allows for
expressing the sum of the contributions from any set of states by means of just one formula
\cite{been.0,furus}. In general, getting a closed-form formula for the integral expressions
one obtains in this way is quite hard, even in the simple case of a superconducting quantum 
point contact (``short junction limit'') \cite{been.0,furus}. On the other hand, in the following we 
show that the formulas for the dc Josephson current greatly simplify in the complementary, long
junction, limit.  As remarked in \cite{bardeen}, in this limit a huge complication arises from
the fact that the net current is ``a small quantity'' that arises from mutual cancellations of large, 
oscillating contributions. In fact, the large oscillations in the function giving the contributions to 
the dc Josephson current from states at a given energy makes it extremely difficult to resort to
a numerical calculation of the total current, even for very short junctions. In order to overcome
such a problem, at the end of this section we will show how, using general mathematical
properties of the $S$-matrix, it is possible to deform the integration path, so to write the
dc Josephson current as just one integral computed over the imaginary axis.  This approach
was originally introduced   in the framework of a Green's function approach, 
by Ishii \cite{ishi}, who used it to show how, on carefully carrying out the sum over
scattering states, a previous result obtained by Kulik \cite{kulik} should be corrected, thus
eventually getting a sawtooth-like dc Josephson current, in the case in which there are no normal  
backscattering processes at the S-N interfaces.   Here,  we employ an adapted version
of the $S$-matrix approach, discussed in \cite{been.0,furus}
for a superconducting quantum point  contact, which allows us to explicitly
compute $I [ \chi ; T ]$ for a generic long SNS matrix and to show that it depends on
data at the Fermi level only. When $T \to 0$ and the single-particle backscattering
at the Fermi level is purely-Andreev-like at both S-N interfaces, we
recover Ishii's sawtooth-like dc Josephson current.   In particular, we consider a general SNS model in the 
non-interacting, BCS approximation with a gap function $\Delta (x)$ of magnitude 
$\Delta$ at $|x|\to \infty$ and $0$ in the central region, $0<x<\ell$. We also include a normal potential, $V(x)$ which vanishes 
at $|x|\to \infty$.  A $4\times 4$ transmission
matrix, $M$ may be defined which relates the asymptotic wave-function in the S regions at $x\to \pm \infty$, 
$\vec A^+=M\vec A^-$ with Bogoliubov-DeGennes wave-function obeying, 
respectively:
\begin{eqnarray}
&~& \left[ \begin{array}{c} u ( x )  \\ v ( x )  \end{array} \right]\to 
\left[ \begin{array}{c}  \cos (\Psi /2)\\ -e^{- i\chi /2}\sin (\Psi /2) \end{array}\right]  
\left[ A_1^- e^{i\beta_p x}+A_2^- e^{-i\beta_p x}\right]
 \nonumber \\   
 &+& \left[ \begin{array}{c} -e^{  i\chi /2}\sin (\Psi /2) \\  \cos (\Psi /2)
   \end{array}\right] \left[A_3^- e^{- i\beta_hx}+A_4^- e^{i\beta_hx}\right] 
\; ,   
   \label{as.1}
\end{eqnarray}
\noindent
for $x \to - \infty$, and
\begin{eqnarray}
&~& \left[ \begin{array}{c} u ( x )  \\ v ( x )  \end{array} \right]\to 
\left[ \begin{array}{c}  \cos (\Psi /2)\\ -e^{  i\chi /2}\sin (\Psi /2) \end{array}\right]  
\left[ A_1^+ e^{i\beta_p (x - \ell) }+A_2^+ e^{-i\beta_p (x - \ell)}\right]
 \nonumber \\   
 &+& \left[ \begin{array}{c} -e^{  - i\chi /2}\sin (\Psi /2) \\  \cos (\Psi /2)
   \end{array}\right] \left[A_3^+ e^{- i\beta_h ( x - \ell) }+A_4^+ e^{i\beta_h ( x - \ell)}\right] 
\; ,   
   \label{as.2}
\end{eqnarray}
\noindent
for $x \to +  \infty$. The particle and hole momenta, for energy $E$, are 
$\beta_{p/h}^2 = 2 m_S \{ \mu \pm ( E^2 - \Delta^2 )^\frac{1}{2} \}$, with $m_S$ the electron effective mass in the S regions,
  $\mu$ is the chemical potential,
and $\Psi \equiv -\arcsin (\Delta /E)$.
(For simplicity, we take an energy-independent gap, $\Delta$, but our results can be extended to 
more realistic models.) 
Eqs.(\ref{as.1},\ref{as.2}) apply also to the 
ABS regime, in which $E^2 - \Delta^2 < 0$. In this case, the
phases of the arguments of the complex square root functions
are always chosen so that $\hbox{Im} ( \beta_p ) \geq  0$ and
$\hbox{Im} ( \beta_h ) \leq  0$ \cite{been.0}. 
The $S$-matrix, which expresses outgoing waves ($A_1^+$, $A_3^+$, $A_2^-$, $A_4^-$) in terms of incoming waves 
can be expressed in terms of $M$. (An alternative way of writing ${\rm det} [S]$ has been introduced in \cite{been.0} 
where it was shown that, in the so-called ``Andreev approximation'', discussed below in
detail, one gets ${\rm det} [ S] = {\rm det} [ {\bf I} - \alpha^2_E r_A^\dagger s_0 ( E ) 
r_A s_0^\dagger (-E)]$,  with $\alpha_E=\exp [ - i {\rm arccos} ( E  / \Delta )]$, 
$r_A = \left[ \begin{array}{cc} e^{ \frac{i}{2} \chi} & 0 \\ 0 &  e^{ -\frac{i}{2} \chi} 
\end{array} \right]$, and $s_0 ( E )$ is the (2$\times$2) scattering matrix for the 
whole system, in the limit of normal leads - $\Delta \to 0$.) In particular we find it convenient to express 
${\rm det} [ S ]$ as a ratio:
\beq {\rm det} [ S ]=\frac{M_{1,1} M_{3,3} - M_{1,3} M_{3,1} }{M_{2,2} M_{4.4} - 
M_{2,4} M_{4,2} }={{\cal F} ( E ; \chi ) \over {\cal G}( E ; \chi ) }\label{mma.1}\ee   
where ${\cal F} ( E ; \chi )$  and ${\cal G} ( E ; \chi )$
may be  regarded as functions of $E$ in the complex $E$-plane
  (see \ref{derivat} for the explicit derivation of Eq.(\ref{mma.1})). 
 We chose  them to obey several convenient properties, which are
crucial for our derivation (and appear to be generally met in physically
relevant models):\\
{\it i)} They are always finite   for   finite $E$. This can be easily achieved by 
shifting poles of ${\cal G}$ into zeroes of ${\cal F}$ and vice versa; \\
{\it ii)} They have no common zeroes. [Possible common zeroes (e.g. $E_0$), could 
always be cancelled by a redefinition: ${\cal F} ( E ; \chi )\to 
{\cal F} ( E ; \chi )/(E-E_0)$, ${\cal G} ( E ; \chi )\to {\cal G} ( E ; \chi )/(E-E_0)$, 
without changing Eq.  (\ref{mma.1})].\\
{\it iii)} ${\cal F} ( E ; \chi )={\cal G}^* ( E ; \chi )$. 
Here this equation refers to complex conjugating {\it the function} without complex 
conjugating its argument, $E$. This condition is consistent with 
the requirement that $ | {\rm det} [ S ] | = 1$ for scattering states.\\
{\it iv)}  ${\cal G} ( E ; \chi )$ can be defined to have  branch cuts along the real $E$-axis,
corresponding to the nonzero density of scattering states in the leads. This is due to the fact that
${\cal G} ( E ; \chi )$ depends on $E$ via $\beta_p $ and $\beta_h$ and that they
become double-valued functions of $E$, for $| E | > \Delta$.\\
{\it v)} $\partial_\chi \ln {\cal G} ( E ; \chi )   $ vanishes rapidly at $|E|\to \infty$
 along any ray not parallel to the real axis. This condition is crucial to
allow for conveniently deforming the integration path in the energy plane, 
when computing $I^{(0)} [ \chi ]$.\\
{\it vi)} ${\cal G} ( E ; \chi )$  is real in the bound state region: the real axis with $- \Delta 
\leq E \leq \Delta$.

These conditions appear to determine ${\cal F} ( E ; \chi )$
 and ${\cal G} ( E ; \chi )$   uniquely except for an 
overall multiplicative constant factor. Moreover, they imply that
zeroes of ${\cal G} ( E ; \chi )$ correspond to poles of
${\rm det} [ S ]$. These conditions imply that
   there are no poles of ${\rm det} [ S ]$ off the real axis. 
We are actually dealing with a 2-sheeted Riemann surface, due to the branch 
cuts. We may regard ${\cal F} ( E ; \chi )$ as being ${\cal G} ( E ; \chi )$ on the second sheet of the
Riemann surface.  With the definition we gave of $\beta_p (E) , \beta_h ( E ) $,
the zeroes of ${\cal G} ( E ; \chi )$,  corresponding to 
poles of ${\rm det} [ S ]$, occur either on the real axis, or else off-axis on the second sheet of the 
Riemann surface. This property  of ${\rm det} [ S ]$  follows from general principles. Since the
$S$-matrix  can be derived from the retarded Green's function it 
should have no singularities in the upper half plane. Since 
\be {\cal G} ( E ; \chi )= {\cal G}^* ( E ; \chi )\label{G*}\ee
 in the BS region, we can use 
the Schwartz reflection principle to define its unique analytic continuation to the entire first sheet of the
Riemann surface, where 
it obeys Eq.  (\ref{G*}). 
Thus if ${\cal G} ( E ; \chi )$ had a zero in the lower half-plane, at $E_0$, it would have to 
have a twin at energy $E_0^*$ in the upper half-plane. This would violate this basic property of $S$ 
telling us that no such zeroes exist. (Notice that the functions ${\cal F} ( E ; \chi )$
 and ${\cal G} ( E ; \chi )$ can be equally well defined in different models, such
as the one describing the Josephson current in ballistic superconductor-graphene systems
\cite{hagy}.)

The ABS's correspond to poles of the $S$-matrix and therefore to the zeroes of ${\cal G}(E)$. This allows 
us to write the contribution to the ground state energy from ABS's as:
\be 
E^{(0)}_{ABS}=-{1\over 2\pi i}\oint_{\Gamma_{ABS}}dE\ln {\cal G} (E;\chi )\label{EABS}\ee
where the contour $\Gamma_{ABS}$ in the complex energy plane surrounds the negative energy ABS's. 
In order to compute the contribution to the ground state energy arising from
the SS's,  $E_{\rm SS}^{(0)}$, we   put the system in a large box $- \frac{L}{2} < x < \ell + \frac{L}{2}$, 
requiring $u(x)$ and $v(x)$  to obey vanishing boundary conditions. This gives an equation of the form:
\be \sum_{n=\pm 1,m=\pm 1}\mu_{n,m}e^{iL (n\beta_p+m\beta_h)}+\mu_{0,0}=0
\:\: , \label{mudef}
\label{bce}\ee
where the coefficients $\mu_{n,m}$ depend on the transmission matrix.  In 
\ref{derivat}, we outline the derivation of Eq.(\ref{mudef}) and, in particular, of
the coefficients $\mu_{n,m}$ in terms of the transmission matrix $M$.  Here we notice that,  defining
\be \zeta \equiv e^{iL(\beta_h-\beta_p)}\:\:\: , \: \:
\eta \equiv e^{iL(\beta _p+\beta_h)}\ee
 and multiplying Eq.  (\ref{bce}) by $\zeta$ 
gives a quadratic equation in $\zeta$ where only the term linear in $\zeta$ depends on $\eta$. 
It then follows that the product of the two roots is independent of $\eta$ and is given by:
\be \prod_{a=1}^2\exp iL(\beta_h^a-\beta_p^a)={\mu_{1,-1}\over \mu_{-1,1}}=\det  [S]\label{nsns.5c}.\ee
Eq. (\ref{nsns.5c}) then implies 

\beq
{\cal G} ( E ; \chi ) =  \prod_{a=1}^2\exp \left[ - \frac{i}{2} L(\beta_h^a-\beta_p^a) \right]
\:\:\:\: . 
\label{one.more}
\eneq
\noindent
At energies for which both $\beta_p$ and $\beta_h$ are real, in particular for
 $- \sqrt{\mu^2 + \Delta^2} \leq E < - \Delta$, 
we may write the solutions of Eq. (\ref{nsns.5c}) in the form:
\beq
\beta_p^a = \frac{\pi m_p}{L} + \frac{\sigma_p^a}{L} \;\;\; , \;\;
\beta_h^a = \frac{\pi m_h}{L} + \frac{\sigma_h^a}{L}
\:\:\:\: 
\label{mmm.1}
\eneq
with 
\be (\beta_p^a)^2/2m_S-\mu = \mu -(\beta_h^a)^2/2m_S.\label{bph}\ee
\noindent
Here $m_p , m_h$ are integers and the phase shifts 
obey $0 \leq \sigma_p^a , \sigma_h^a \leq  \pi$.
Thus the phase shifts obey
\beq
{\cal G} ( E ; \chi ) = \prod_a e^{ \frac{i}{2} [ \sigma_p^a - \sigma_h^a ] } 
\:\:\:\: . 
\label{mmmm.a1}
\eneq
\noindent
  By adapting the derivation of \cite{ACZ} to the continuum model, we
now derive the contribution to the total groundstate energy
arising from states with energy between $- \sqrt{\mu^2 + \Delta^2}$ and $- \Delta$.
From the definition of $\beta_p^a , \beta_h^a$ in Eq.(\ref{mmm.1}), as $L \to \infty$,  one obtains
$(\beta_p)^2 / 2 m_S - \mu = \mu - ( \beta_h )^2 / 2 m_S$ and 
$\beta_p \sigma_p^a = - \beta_h \sigma_h^a$, with $\beta_{p/h} = \pi m_{p/h} /L$. 
Defining $\beta_{p; l(u)} , \beta_{h; l (u)}$ to be the values of $\beta_p , \beta_h$ corresponding to
$- \sqrt{\mu^2 + \Delta^2}$ and to $- \Delta$, respectively, one then finds that the total groundstate energy
arising from states with energy between $- \sqrt{\mu^2 + \Delta^2}$ and $- \Delta$,
$E^{1}_{\rm SS}$, may be either written as $E^{1}_{\rm SS} = {\cal E}_{\rm SS; p} + 
\frac{1}{\pi m_S} \: \int_{\beta_{p;l}}^{\beta_{p;u}} \: d \beta_p \: \beta_p \sigma_p^a $, or 
as $E^{1}_{\rm SS} = {\cal E}_{\rm SS; h} -
\frac{1}{\pi m_S} \: \int_{\beta_{h;l}}^{\beta_{h;u}} \: d \beta_h \: \beta_h \sigma_h^a $, with
${\cal E}_{\rm SS; p} , {\cal E}_{\rm SS; h} $ being independent of $\chi$. Taking the
mean of the two equivalent expressions for $E^{1}_{\rm SS}$, we may then write :  
\begin{eqnarray} 
E^{1}_{\rm SS}  &=&  {\cal E}^1_{\rm SS} + \frac{1}{2 \pi } \: \int_{-\sqrt{\mu^2 + \Delta^2}}^{-\Delta} \: d E \: 
\sum_a [  \sigma_p^a - \sigma_h^a ] \nonumber \\
&=& {\cal E}^1_{\rm SS} - 
\frac{1}{2\pi i } \: \int_{-\sqrt{\mu^2 + \Delta^2} }^{-\Delta} \: d E \: \ln \left[ \frac{ {\cal G}^* ( E ; \chi )}{
{\cal G}  ( E ; \chi ) } \right]  
\:\:\:\: ,
\label{mmm.4a}
\end{eqnarray}
\noindent
with ${\cal E}_{\rm SS}^1$ being independent of $\chi$. Remarkably, Eq. (\ref{mmm.4a}) 
can be readily extended to $E < - \sqrt{\mu^2 + \Delta^2}$, in which case
one obtains ${\cal G} ( E ; \chi ) =  
\prod_a e^{ - i  \sigma_h^a  }$. A procedure similar  to the one leading
to Eq. (\ref{mmm.4a}) yields the
contribution to the total groundstate energy
arising from states with energy $E < - \sqrt{\mu^2 + \Delta^2}$, 
$E^2_{\rm SS}$, which is given by

\begin{eqnarray}
E^{2}_{\rm SS} &= &{\cal E}^2_{\rm SS} -  \frac{1}{\pi } \: \int^{-\sqrt{\mu^2 + \Delta^2}}_{-\infty} \: d E \: 
\sum_a  \sigma_h^a  \nonumber \\   
&=& {\cal E}^2_{\rm SS} - 
\frac{1}{2\pi i } \:  \int^{-\sqrt{\mu^2 + \Delta^2}}_{-\infty}  \: d E \: \ln \left[ \frac{ {\cal G}^* ( E ; \chi )}{
{\cal G}  ( E ; \chi ) } \right]  
\:\:\:\: ,
\label{mmm.4b}
\end{eqnarray}
\noindent
with  ${\cal E}^2_{\rm SS}$ being independent of $\chi$. Adding Eqs. (\ref{mmm.4a},\ref{mmm.4b}), 
 we obtain the total contribution to the ground 
state energy from scattering states in the form:
\be 
E^{(0)}_{SS}=\epsilon^0_{SS}-{1\over 2\pi i}\int_{-\infty}^{-\Delta}dE\ln \left[{{\cal G}^*(E;\chi )\over 
{\cal G}(E;\chi )}\right]\label{ESS}\ee
where $\epsilon^0_{SS}$ is independent of $\chi$. 
Using Eq.  (\ref{G*}), the second term in Eq.  (\ref{ESS}) can be also be written as a contour integral in the complex energy plane, 
 like Eq.  (\ref{EABS}), 
with the contour now running on both sides of the branch cut in ${\cal G}$ along the real $E$-axis 
from $-\infty$ to $-\Delta$. 
These two terms can be combined, allowing us to write a simple unified formula 
for the zero-temperature Josephson current:
\beq
I^{(0)} [ \chi ] = - \frac{2 e}{2 \pi i } \: \int_{\Gamma } \: d E \: 
\partial_\chi \{ \ln {\cal G} ( E ; \chi ) \}  
\label{com.4}
\eneq
where the contour $\Gamma$ runs infinitesimally above and below the negative $E$ axis. 

\begin{figure}
\centering \includegraphics*[width=1.00\linewidth]{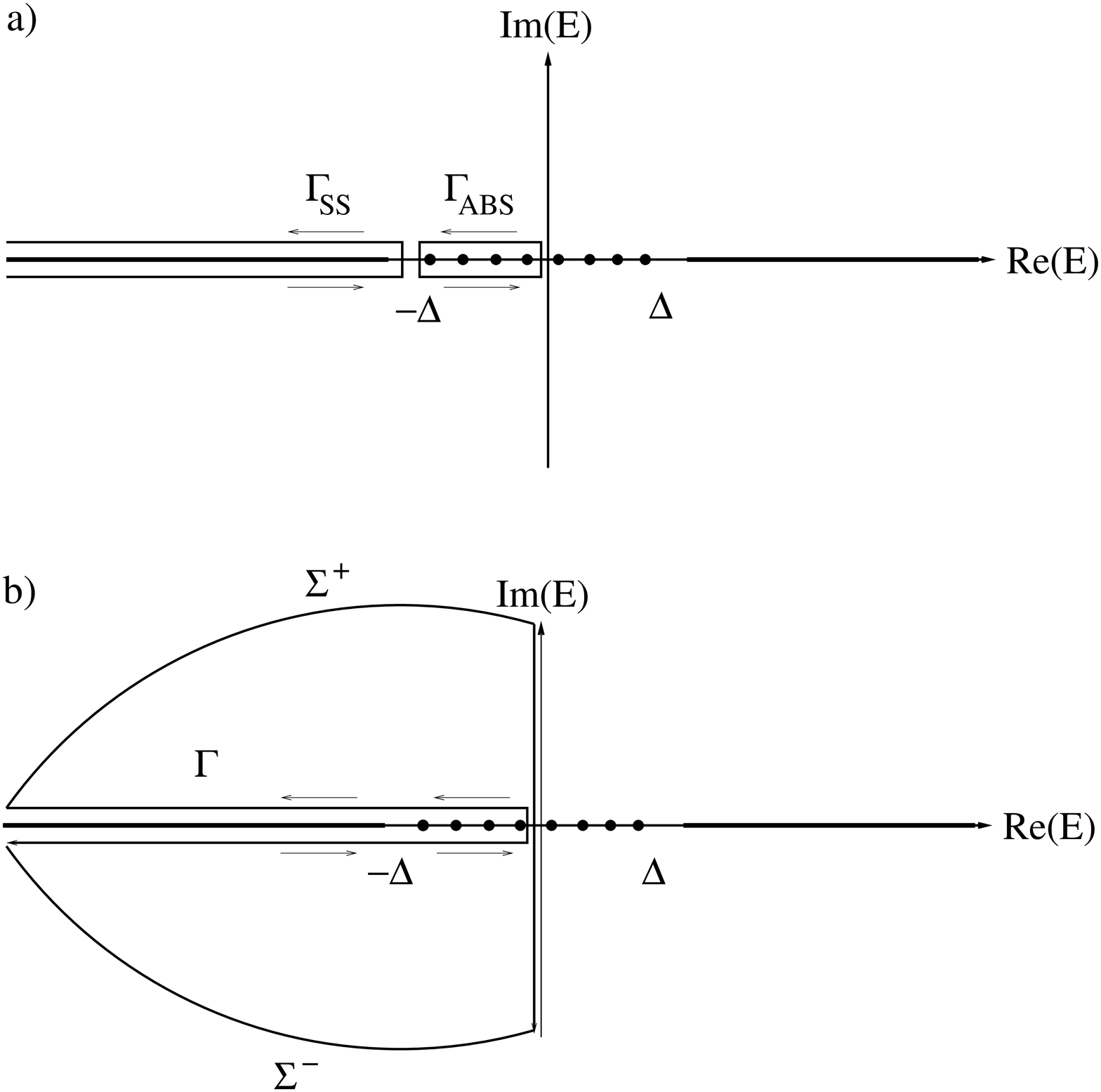}
\caption[]{Sketch of the deformation of the integration path
$\Gamma$ used in Eq. (\ref{com.4}) to get to Eq. (\ref{com.4.az}):
\\
{\bf a)} The integration path $\Gamma$ sketched as $\Gamma = \Gamma_{\rm ABS} \cup \Gamma_{\rm SS}$, with
$\Gamma_{\rm ABS}$ running around poles corresponding to ABS's and $\Gamma_{\rm SS}$ going around
the energy interval corresponding to negative-energy SS's; \\
{\bf b)} Adding the arcs $\Sigma^+ , \Sigma^-$, whose contribution to the integral is
zero as their radius is sent to $\infty $ (see text for the discussion), allows
for trading the integral over $\Gamma$ for an integral over the imaginary axis.}
\label{prl1_f}
\end{figure}

Eq. (\ref{com.4}) is exact and, in principle, as long as  ${\cal G} ( E ; \chi )$ is
known, it may be used to compute $I^{(0)} [ \chi ]$ for any values of 
the system parameters. However, in general it is of no great usefulness for
practical purposes as, typically, when $E$ lies on the real axis,
$ {\cal G} ( E : \chi ) $ turns out to be a rapidly oscillating function of
$E$, which makes it quite hard to figure out reliable approximations in
possibly relevant regimes (such as, for instance, the ``long junction'' limit).
In addition, oscillations  also make any attempt to numerically estimate  $I^{(0)} [ \chi ] $
fail, except possibly in some very specific cases, such as the  short junction
limit. A way to greatly improve the convergence properties of the integral
in Eq. (\ref{com.4}) is    to deform the integration path $\Gamma$ by using 
the fact that, on the physical Riemann sheet,   ${\cal G} ( E ; \chi )$ has no
zeroes off the real axis and that $\partial_\chi \ln {\cal G} ( E ; \chi )$ vanishes 
 rapidly at $|E|\to \infty$. Thus, to trade Eq. (\ref{com.4}) for a more
tractable formula, we use the fact that, due to the properties of ${\cal G} (E; \chi )$ 
discussed above, $\Gamma$  can be deformed 
 into a single line running along the imaginary $E$-axis from $-\infty$ to $\infty$.
[See Fig. (\ref{prl1_f}).] As a result, one eventually obtains the
general formula

\beq
I^{(0)} [ \chi ] =  \frac{2 e}{2 \pi   } \: \int_{- \infty }^\infty \: d \omega \: 
\partial_\chi \{ \ln {\cal G} (i  \omega : \chi ) \}  
\:\:\:\: . 
\label{com.4.az}
\eneq
\noindent
Eq. (\ref{com.4.az}) is particularly amenable for explicitly computing
$I^{(0)} [ \chi ]$ for at least two  reasons: first of all, integrating
over the imaginary axis greatly improves the convergence properties of 
the integral, as it allows for getting rid of the oscillations
in the integrand functions. Moreover, as we will show in the
explicit examples discussed in the following, it enables us to compute
$I^{(0)} [ \chi ]$ in a systematic expansion in inverse powers of
the length of the junction (that is, of the size $\ell$ of the normal
region C), eventually letting us show that, to leading order
in $\ell^{-1}$, terms in $I_{\rm SS}^{(0)} [ \chi ]$ and $I^{(0)}_{\rm ABS} [ \chi ]$
cancel with each other, so that $I^{(0)} [ \chi ]$
 can be expressed in terms of  ABS's at the Fermi level only.

\section{dc Josephson current in a long SNS junction}
\label{long}

\begin{figure}
\centering \includegraphics*[width=1.00\linewidth]{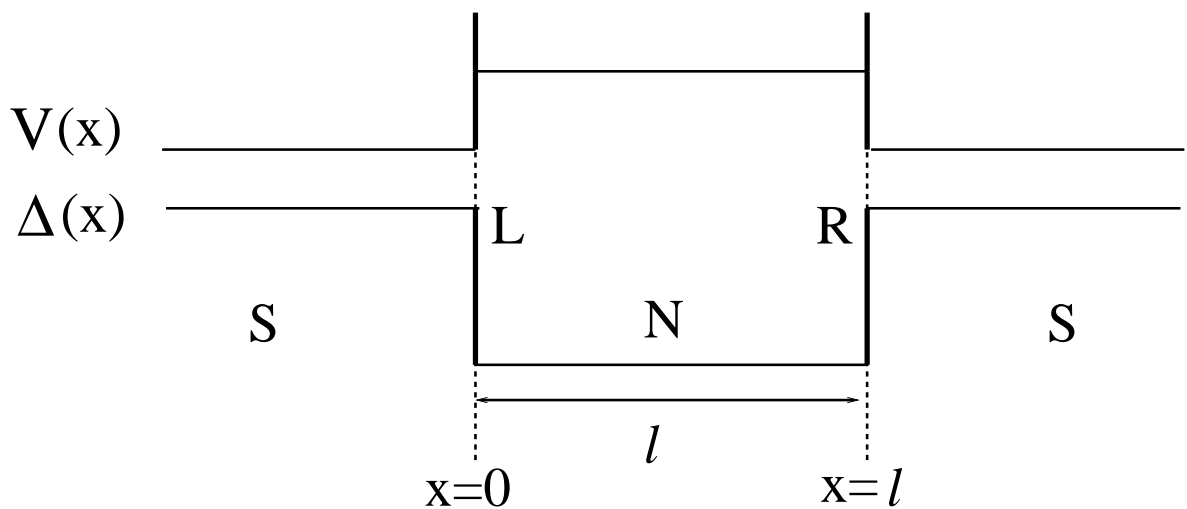}
\caption[]{Sketch of $V(x) , \Delta ( x )$ in the SNS system. The central region C
has length $\ell$ and is defined by $0< x < \ell$. The L- and R-interfaces are 
assumed to be ``sharp'', that is, $V ( x ) , \Delta ( x )  $ and the single-particle
mass are assumed to vary over typical length scales $\ll \ell$. In this case
scattering is localized at the interfaces and is fully encoded
within the particle- and hole- normal- and Andreev- scattering amplitudes at
the interfaces, $\bar{N}_{L/R}^{p/h} ,  \bar{A}_{L/R}^{p/h}$.}
\label{prl0_f}
\end{figure}

We assume that our system is made of two
superconductors at phase difference $\chi$ separated by a long central
normal region C of length $\ell$, defined by $0 < x < \ell$,  as sketched in Fig. (\ref{prl0_f}).
   We assume that the gap function $\Delta (x)$ makes an abrupt transition 
from $\Delta e^{\pm i\chi /2}$ to $0$ in the central region and  include a normal potential energy function $V(x)$ which we assume makes 
an abrupt transition from $0$ in the leads to $V_C$ in the central region. Here ``abrupt'' means rapid on 
the scale of $\ell$.  We assume $V_C<\mu$ so that the central region is metallic.  The mass in C is written as $m$. 
The wave-functions in the central region, far from the interfaces, may be written as

\beq
\left[ \begin{array}{c} u(x) \\ v( x ) \end{array} \right]
= \left[ \begin{array}{c}  C_1\exp (i\alpha_p x)+C_2\exp (-i\alpha_p x)  \\
  C_3\exp (-i\alpha_h x)+C_4\exp (i\alpha_h x) \end{array} \right]
\;\;\;\; , 
\label{new.1}
\eneq
\noindent
with $\alpha_{p/h} = \{ 2m ( \mu -V_C\pm E )\}^\frac{1}{2} $. Thus, we define the transmission matrices for the left 
and right interfaces, $L$ and $R$ by $\vec C=L\vec A^-$, $\vec A^+=R\cdot M^C \vec C$
in terms of which $M = R \cdot M^C \cdot L$, where $M^C$ is the transmission matrix
of C, given by  

\beq
M^C = \left[ \begin{array}{cccc} e^{ i\alpha_p\ell } &0 &0 & 0 \\ 0 & 
e^{ -i\alpha_p\ell } & 0 & 0 \\ 0 & 0 & e^{-i\alpha_h\ell } & 0 \\
0 & 0 & 0 & e^{ i\alpha_h\ell } \end{array} \right]
\:\:\:\: . 
\label{new.2}
\eneq
\noindent
By definition of ${\cal F}$ and ${\cal G}$ in Eq. (\ref{mma.1})
and of the $R$- and $L$-transmission
matrices, one finds that the following explicit
formulas hold:

\begin{eqnarray}
&{\cal F}& ( E ; \chi ) = F_{0,0} (  \chi ; E ) 
 +  F_{1,1} (E)  e^{ i 
[ \alpha_p - \alpha_h ] \ell}  \nonumber \\
 &+&  F_{-1,-1} (E)  e^{ - i 
[ \alpha_p - \alpha_h ] \ell} 
+  F_{1,-1}  (E)  e^{ i 
[ \alpha_p + \alpha_h] \ell} \nonumber \\
&+& F_{-1, 1} (E)   e^{ - i 
[ \alpha_p +\alpha_h ] \ell}\nonumber \\
&{\cal G}& ( E ; \chi ) =G_{0,0} (  \chi ; E ) + G_{1,1}  (E)   e^{ i 
[ \alpha_p - \alpha_h ] \ell} \nonumber \\
&+& G_{-1,-1}  (E)  e^{ - i 
[ \alpha_p - \alpha_h ] \ell} +  G_{1,-1}  (E)  e^{ i 
[ \alpha_p + \alpha_h] \ell} \nonumber \\
&+& G_{-1, 1} (E)   e^{ - i 
[ \alpha_p + \alpha_h] \ell},
\label{rafi.1}
\end{eqnarray}
\noindent
with

\begin{eqnarray}
 F_{0,0} (   \chi ) &=&  -  
( R_{1,1} R_{3,2} - R_{1,2} R_{3,1} )(L_{1,3} L_{2,1}
- L_{1,1} L_{2,3} ) \nonumber \\
&-&     ( R_{1,3} R_{3,4} - R_{1,4} R_{3,3} ) ( L_{3,3} L_{4,1} -
L_{3,1} L_{4,3} ) \nonumber \\
F_{1,1}  &=&    (R_{1,1} R_{3,3} - R_{1,3} R_{3,1} )(L_{1,1}  
L_{3,3}   -  L_{1,3}  L_{3,1} ) \nonumber \\
F_{-1,-1}  &=&  ( R_{1,2} R_{3,4} - R_{1,4} R_{3,2} )( L_{2,1}  L_{4,3} 
-  L_{4,1} L_{2,3} ) \nonumber \\
F_{1,-1}  &=&    ( R_{1,1} R_{3,4} - R_{3,1} R_{1,4} )( L_{1,1}  L_{4,3}  - 
L_{1,3}  L_{4,1}  ) \nonumber \\
F_{-1,1}   &=&   ( R_{1,2} R_{3,3} - R_{3,2} R_{1,3} )( L_{2,1}  L_{3,3}  - 
L_{2,3}  L_{3,1}  )  \; , \nonumber \\  
\label{m.5a}
\end{eqnarray}
\noindent
and

\begin{eqnarray}
 G_{0,0} (  \chi ) &=&    -    ( R_{2,1} R_{4,2} - R_{2,2} R_{4,1} )(L_{1,4} L_{2,2}
- L_{1,2} L_{2,4} ) \nonumber \\
&-&   ( R_{2,3} R_{4,4} - R_{2,4} R_{4,3} ) ( L_{3,4} L_{4,2} -
L_{3,2} L_{4,4} ) \nonumber \\
G_{1,1}  &=&   (R_{2,1} R_{4,3} - R_{2,3} R_{4,1} )(  L_{1,2}  
L_{3,4}  -  L_{3,2}  L_{1,4} ) \nonumber \\
G_{-1,-1}   &=&   ( R_{2,2} R_{4,4} - R_{2,4} R_{4,2} )(
L_{2,2}  L_{4,4}   - L_{2,4} L_{4,2}    ) \nonumber \\
G_{1,-1}   &=&    ( R_{2,1} R_{4,4} - R_{4,1} R_{2,4} )( L_{1,2}  L_{4,4}  - 
 L_{1,4}  L_{4,2} ) \nonumber \\
G_{-1,1}   &=&    ( R_{2,2} R_{4,3} - R_{4,2} R_{2,3} )( L_{2,2}  L_{3,4}  - 
L_{3,2}  L_{2,4}  )    \; , \nonumber \\ 
\label{m.6a}
\end{eqnarray}
\noindent
and the explicit dependence upon $\chi$ set according to the definition
of the $R$- and $L$-matrices. Using general properties of the transmission
matrices, arising from the continuity equation for probability
current, it is not difficult to show that Eqs. (\ref{m.5a},\ref{m.6a}) 
are consistent with the identity   ${\cal F} ( E ; \chi ) = {\cal G}^* ( E ; \chi)$, 
that ${\cal G} ( E ; \chi)$
is real for real $E$ and $- \Delta \leq E \leq \Delta$, and that 
the branch cuts of ${\cal G} ( E ; \chi )$ lie on the
real axis, from $E \to - \infty$ to $E = - \Delta$ and from $E = \Delta$ to
$E \to \infty$.  From Eqs. (\ref{rafi.1}), one then sees that 
 
\beq
\partial_\chi \ln {\cal G} ( E ; \chi ) = \frac{\partial_\chi G_{0,0} ( E ; \chi ) }{
{\cal G} ( E ; \chi ) }
\:\:\:\: . 
\label{obser.1}
\eneq
\noindent
Because, as  $E$ goes to infinity along any ray not parallel to the real axis, either the imaginary part of 
$\alpha_p$, or the imaginary part of $\alpha_h$, goes to
$- \infty$, from Eq. (\ref{obser.1}) we find that 
$\partial_\chi \ln {\cal G} ( E ; \chi )   $ exponentially vanishes as 
$|E|\to \infty$ off the real axis. Using Eq. (\ref{com.4.az}) for
the current and taking into account that, for large $\ell$, 
 contributions to the integral   with $| \omega |  \geq V$ are strongly 
suppressed, we may approximate $\alpha_p$ and $\alpha_h$ as

\begin{eqnarray}
\alpha_p & \approx& \alpha_F +  i \omega \sqrt{\frac{ m}{2 (\mu -V_C)}}
\nonumber \\
\alpha_h  &\approx&  \alpha_F -  i \omega \sqrt{\frac{ m}{2 (\mu -V_C)}}
\:\:\:\: , 
\label{com.6}
\end{eqnarray}
\noindent
with $\alpha_F = \sqrt{2 m ( \mu - V_C )}$.
Using Eqs. (\ref{com.6}), the zero-temperature Josephson current 
to leading order in $\ell^{-1}$ is then given by

\begin{eqnarray}
 I^{(0)} [ \chi ]  &=&   -  \frac{e}{\pi \ell} \sqrt{\frac{\mu - V_C}{2 m } } \:  \: \int_{-\infty}^\infty \: d z \: 
 \partial_\chi \bar{G}_{0,0} ( \chi ) \{ \bar{G}_{1, 1} e^{ - z}  + 
\bar{G}_{- 1,-1}e^{  z} \nonumber \\
 &+& \bar{G}_{1,- 1}  e^{ 2 i \alpha_F \ell}  
 +  \bar{G}_{- 1, 1} e^{ - 2 i \alpha_F \ell}   + \bar{G}_{0,0} ( \chi ) \}^{-1} 
\:\:\:\: , 
\label{com.7}
\end{eqnarray}
\noindent
with the coefficients $\bar{G}_{a,b}$ being defined as the coefficients 
$G_{a,b}$ evaluated at $\omega = 0$, that is, setting $\alpha_p =  \alpha_h = \alpha_F$, 
$\beta_p =  \beta_h^* = \{ 2 m_S [ \mu + i \Delta ] \}^\frac{1}{2} $. Computing the integral
in Eq. (\ref{com.7}), one eventually finds out

\beq
 I^{(0)} [ \chi ] =      \frac{e v_F}{ \pi \ell} \:   \: \partial_\chi
\ln^2 \left[ \frac{u_+ ( \chi )}{u_- ( \chi ) } \right]
\:\:\:\: ,
\label{com.9}
\eneq
\noindent
with $v_F = \alpha_F / m$ and    $u_\pm ( \chi )$ being the 
roots of the second-degree equation

\beq
\bar{G}_{-1,-1} u^2 + [ \bar{G}_{1,- 1} e^{ 2 i \alpha_F \ell} +  \bar{G}_{  - 1,1} e^{ -2 i \alpha_F \ell} 
+\bar{G}_{0,0} ( \chi ) ] u + \bar{G}_{1,  1} = 0 
\;\;\;\; . 
\label{com.8}
\eneq
\noindent
Because of particle-hole symmetry at the Fermi level, 
one finds  $\bar{G}_{-1,-1}  = \bar{G}_{1,  1}$, which implies the
identity $u_+ ( \chi ) u_- ( \chi ) =1$ that we used  to derive Eq. (\ref{com.9}).
In order to prove that Eq. (\ref{com.9}) yields Eq. (\ref{main}), 
we have to rewrite Eq. (\ref{com.8}) in terms
of the normal- and Andreev-scattering amplitudes at the
Fermi level. To do so, we relate the scattering amplitudes at
both interfaces to the $R$- and $L$-matrix elements. This
may be readily done starting from the definition of
the normal- and Andreev-scattering amplitudes. The
result is

 \begin{eqnarray}
 N_R^p  ( E )  &=& \frac{R_{2,4} R_{4,1} - R_{2,1} R_{4,4}}{R_{2,2} R_{4,4} - R_{2,4} R_{4,2}}   \nonumber \\ 
 A_R^p  ( E )  &=&  \frac{R_{2,1} R_{4,2} - R_{2,2} R_{4,1}}{R_{2,2} R_{4,4} - R_{2,4} R_{4,2}} \nonumber \\
 N_R^h  ( E )  &=& \frac{R_{2,3} R_{4,2} - R_{2,2} R_{4,3}}{R_{2,2} R_{4,4} - R_{2,4} R_{4,2}}   \nonumber \\  
 A_R^h  ( E )  &=&  \frac{R_{2,4} R_{4,3} - R_{2,3} R_{4,4}}{R_{2,2} R_{4,4} - R_{2,4} R_{4,2}} 
\:\:\:\: ,
\label{amp.12}
\end{eqnarray}
\noindent
and
  
 \begin{eqnarray}
 N_L^p  ( E )  &=& \frac{L_{1,2} L_{4,4} - L_{1,4} L_{4,2}}{L_{2,2} L_{4,4} - L_{2,4} L_{4,2}}   \nonumber \\ 
 A_L^p  ( E )  &=&  \frac{L_{3,2} L_{4,4} - L_{3,4} L_{4,2}}{L_{2,2} L_{4,4} - L_{2,4} L_{4,2}}  \nonumber \\
 N_L^h  ( E )  &=& \frac{L_{2,2} L_{3,4} - L_{2,4} L_{3,2}}{L_{2,2} L_{4,4} - L_{2,4} L_{4,2}}    \nonumber \\ 
 A_L^h  ( E )  &=&  \frac{L_{1,4} L_{2,2} - L_{1,2} L_{2,4}}{L_{2,2} L_{4,4} - L_{2,4} L_{4,2}}  
\:\:\:\: .
\label{ampe.12}
\end{eqnarray}
\noindent
Using Eqs. (\ref{amp.12},\ref{ampe.12}) specified at $E=0$, we
may then rewrite   Eq. (\ref{com.8}) as

\beq
u^2 + 1 - \{   \bar{N}_R^p \bar{N}_L^p e^{2 i \alpha_F \ell}  
+   \bar{A}_R^p \bar{A}_L^h    + {\rm c.c.}   \} u = 0
\:\:\:\: ,
\label{com.8bis}
\eneq
\noindent
with, as specified below, the overbar meaning that  all 
the scattering amplitudes in Eq. (\ref{com.8bis}) are evaluated at $E=0$.
By definition, one has that $\bar{A}_R^p ,\bar{A}_L^h \propto e^{ \frac{i}{2} \chi}$, 
where $\propto$ stays for factors that are independent of $\chi$. 
As a result,  $\bar{A}_R^p \bar{A}_L^h \propto e^{ i \chi}$ and, 
on setting $u_\pm ( \chi ) = e^{ \pm i \vartheta ( \chi)}$, we
obtain 

\beq
\vartheta ( \chi ) = {\rm arccos} 
\{  \hbox{Re} [ \bar{N}_R^p \bar{N}_L^p e^{ 2 i \alpha_F \ell}  
+  \bar{A}_R^p \bar{A}_L^h    ] \}
\;\;\;\; , 
\label{novel.1}
\eneq
\noindent
which finally gives our main result in Eq. (\ref{main}).

Our result makes it feasible to compute the dc Josephson current
in several models of physical interest. We are now going to analyse two of
them, as  examples, in the next section.

\section{Explicit calculation of the dc Josephson current in models of physical interest} 
\label{explicit}

To illustrate the effectiveness of our result, we now compute $I^{(0)} [ \chi ]$ in
two models of physical interest, by also showing how  results previously obtained in
the literature for specific values of the system parameters may be straightforwardly recovered 
within our approach. 

\begin{figure}
\centering \includegraphics*[width=0.80\linewidth]{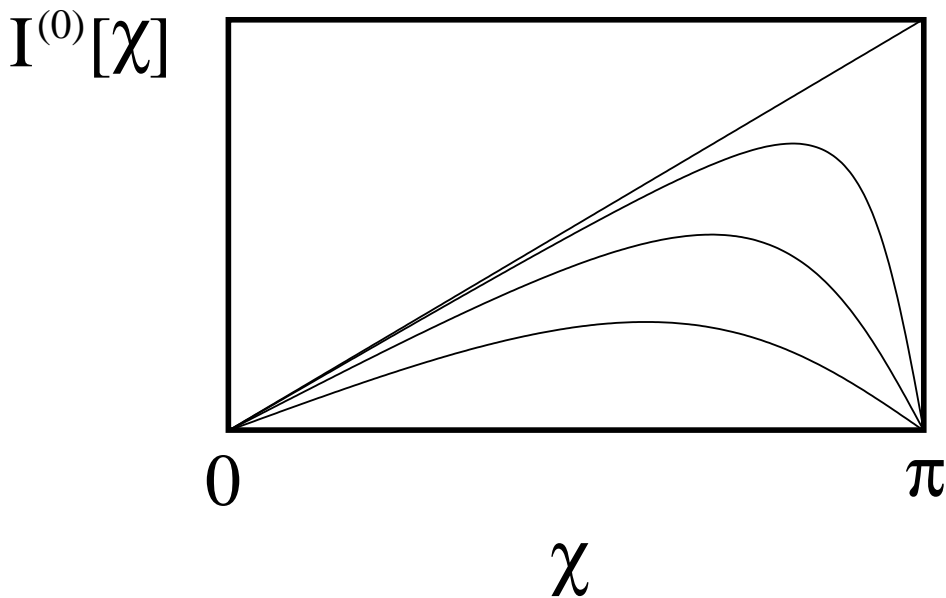}
\caption[]{$I^{(0)} [ \chi]$ {\it vs.} $\chi$ for the two-interface BTK model for different values of
the system parameters, with $\vartheta ( \chi )$ computed with Eq. (\ref{r.test.3}). The 
parameters have been chosen so that $\alpha_F = \hbox{Re} \beta_F $ (arbitrary units),
$\hbox{Im} \beta_F / \hbox{Re} \beta_F  = 0.77  $, $\alpha_F \ell = 13 \pi$, 
while $Z$ is varied, so as to change the ratio between the normal
and the Andreev reflection coefficients. In particular, from bottom to top we set $Z / \hbox{Im} \beta_F =0,0.5 , 
0.8 , 1$. Notice that Ishii's sawtooth behavior for $I^{(0)} [ \chi ]$ is recovered only when $Z$ 
is fine-tuned to be equal to $\hbox{Im} \beta_F$. On the other hand, for $Z=0$, the mismatch between the 
Fermi momenta in the central region and in the superconducting leads (due to $\hbox{Im} \beta_F$ 
being different from zero) yields a nonzero normal reflection coefficient, as evidenced 
by the sinusoidal dependence of $I^{(0)} [ \chi ]$ on  $\chi$.}
\label{lat1_bis}
\end{figure}

As a first example, we consider  a model system whose
S-N interfaces may be thought of as a generalization of the one studied in
\cite{btk}, that is, we assume that $m$ is uniform, $\Delta ( x ) $ abruptly changes
at the S-N interfaces and that the normal potential energy function $V(x)$ is given by
\be V(x)=V_0[\delta (x)+\delta (x-\ell )]+V_C\theta (x)\theta (\ell -x)\ee
where $\delta$ and $\theta$ are the Dirac delta-function and Heavyside step function respectively
\cite{likar}. At the Fermi level, the result for the normal reflection
amplitudes is

\beq
\bar{N}_R^p    =  \bar{N}_L^p   =  - \frac{ ( \beta_F - \alpha_F - i Z )( \beta_F^* + \alpha_F + i Z )}{
\alpha_F^2 + | \beta_F - i Z|^2  } \;\;\;\; , 
\label{r.test.1}
\eneq
\noindent
with $\beta_F = \{ 2 m ( \mu + i \Delta ) \}^\frac{1}{2}$ (that is, 
the momentum $\beta_p$ for $E = 0$), and $Z = 2 m V_0$. 
The Andreev reflection amplitudes at the Fermi level are
given by

\beq
\bar{A}_R^p   = \bar{A}_L^h   = \frac{i e^{ i \frac{\chi}{2}} \alpha_F ( \beta_F + \beta_F^* )}{
\alpha_F^2 + | \beta_F - i Z|^2  } \;\;\;\; . 
\label{r.test.2}
\eneq
\noindent
 Eq.  (\ref{novel.1}) now gives:
\begin{eqnarray}
&~&\vartheta ( \chi ) = {\rm arccos} \biggl\{\hbox{Re} \biggl[ - 
\cos ( \chi ) \left[ \frac{  \alpha_F  ( \beta_F + \beta_F^* ) }{
\alpha_F^2 + | \beta_F - i Z |^2} \right]^2 \nonumber \\
&-&
\left( \frac{ ( \beta_F - \alpha_F - i Z ) ( \beta_F^* + \alpha_F + i Z ) }{
\alpha_F^2 + | \beta_F - i Z |^2} \right)^2 e^{ 2 i \alpha_F \ell } \biggr]\biggr\} 
\label{r.test.3}
\end{eqnarray}
\noindent
where $\alpha_F = \sqrt{2m ( \mu - V_C)}$. 
An interesting point is how Ishii's sawtooth current \cite{ishi} may be recovered
from our result in Eq. (\ref{r.test.3}), taken in an appropriate limit. First of all, 
let us remark that, as shown in Fig. (\ref{lat1_bis}), in order to obtain Ishii's result from the formula
for $\vartheta ( \chi )$ in Eq. (\ref{novel.1}), one has
to take the system in the   limit of perfect Andreev scattering at zero energy, 
$\bar{N}^p_{L/R}=0$, $|\bar{A}_{L/R}^{p/h}|=1$. (On the other hand, notice that one obtains a complete 
suppression of the Josephson current when the Andreev reflection amplitude 
vanishes at either interface.) Thus, we see that 
two conditions must be satisfied for perfect Andreev reflection and hence a sawtooth current:
$Z=\hbox{Im}\beta_F=\sqrt{m(\sqrt{\mu^2+\Delta^2}-\mu )}$, 
$V_C=- Z^2/(2 m)$. These two conditions are readily met if one assumes $V_C = Z=0$
(corresponding to S-N interfaces without barrier normal scattering potential
\cite{btk})  and $\Delta / \mu \approx 0$ (that is, the so-called ``Andreev approximation'',
consisting in assuming no normal scattering at the interfaces at zero energy: in the absence of
barrier potential this is quite a harmless approximation, given the typical values
for $\mu$ and $\Delta$ in an ordinary superconductor). As both approximations are made in
Ishii's derivation, we see how the result of \cite{ishi} may be regarded as just
a special case of our Eq. (\ref{r.test.3}).

\begin{figure}
\centering \includegraphics*[width=0.80\linewidth]{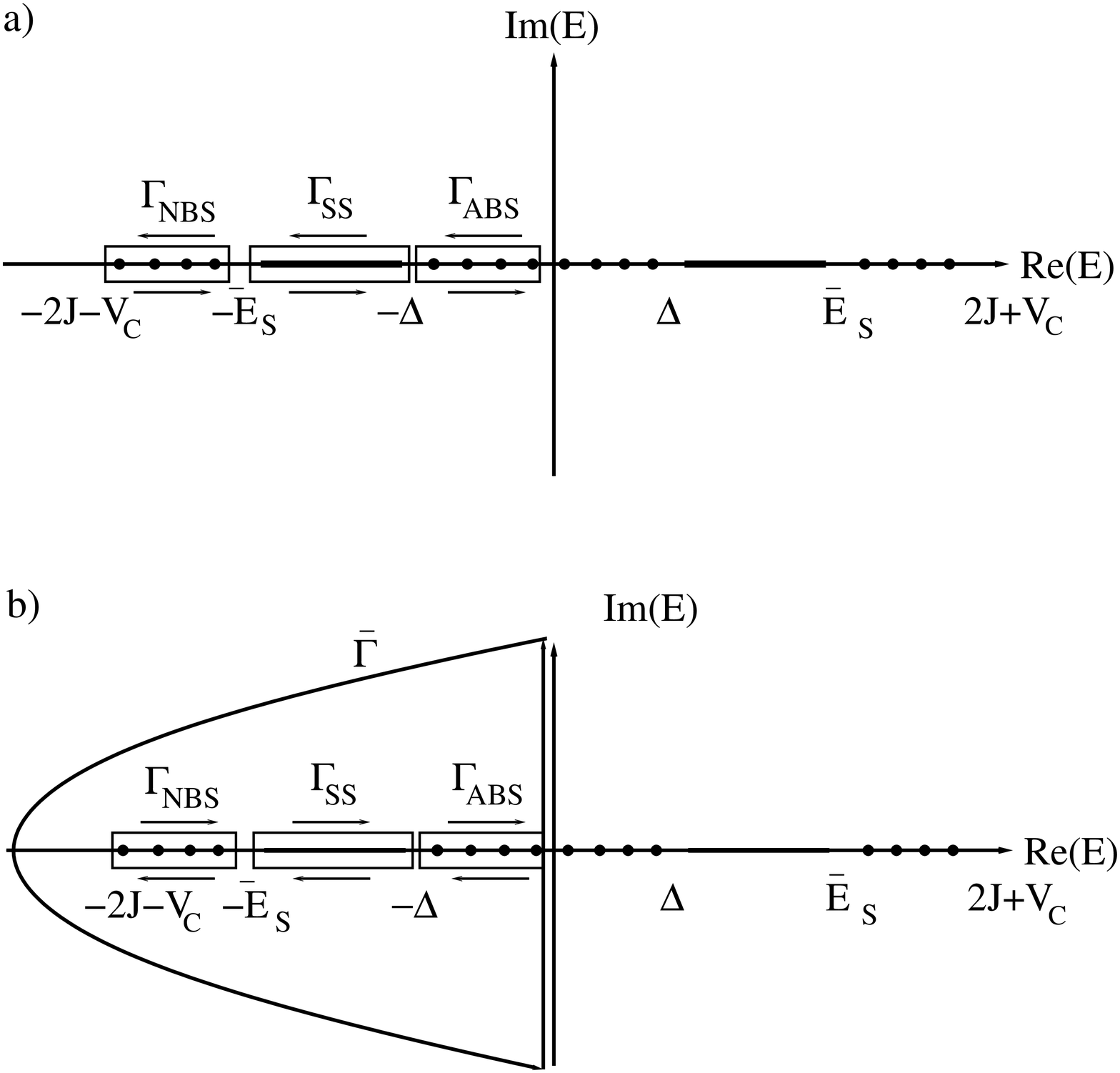}
\caption[]{Integration paths used to compute $I^{(0)} [ \chi ]$
in the lattice model.\\
{\bf a):} Integration paths $\Gamma_{\rm ABS} , \Gamma_{\rm SS} , 
\Gamma_{\rm NBS}$. There is no $\Gamma_{\rm NBS}$ if there
are no NBS's, that is, for $2J + V_C < \bar{E}_S$; \\
{\bf b):} Integration path $\bar{\Gamma}$ used to resort to
integrating over the imaginary axis.}
\label{lat1}
\end{figure}

The second example we consider is related to the fact that, while
in  carrying out our derivation, we mainly referred to
a continuum one-dimensional model of a SNS system just because
of the wide applicability of such a model,  the requirements
on ${\cal F} ( E ; \chi)$ and on ${\cal G} ( E ; \chi )$ we 
made in section \ref{general} are quite general, so, we 
expect our approach to successfully apply to a wide class
of models such as, for instance, the paradigmatic
tight-binding Hamiltonian studied in  \cite{ACZ}. 
In particular, we now   derive the dc Josephson
current for a particular lattice model Hamiltonian for
a central region consisting of $\ell - 1$ sites connected
to two infinite bulk superconductors at phase difference 
$\chi$ \cite{ACZ} .  For such a system, the amplitudes $u_j , v_j$ 
become functions of the lattice site $j$, and the
BDG equations are given by

\begin{eqnarray}
E   u_{ j } &=& - \tau_{ j , j+1} u_{ j +1 } - \tau_{ j , j-1} u_{j - 1 }  - \mu u_j
+ V_j  u_{ j }  + \Delta_j v_{ j } \nonumber \\
E   v_{ j } &=&  \tau_{ j , j+1} v_{ j +1 } + \tau_{ j , j-1} v_{j - 1 } + \mu v_j
- V_j v_{ j }  + \Delta_j^* u_{ j}
.\nonumber \\ &&
\label{csns.1} 
\end{eqnarray}
\noindent
with the lattice hopping amplitudes being given by

\be
\tau_{j , j+1} = \biggl\{ \begin{array}{l} t_S \: {\rm for} \: j \leq -1   \: {\rm and} \:
                        {\rm for} \:j \geq \ell \\
J \: {\rm for} j \in \{ 1 , \ldots , \ell - 2 \} \\
t'' \: {\rm for} \:  j = 0 , \ell - 1
                       \end{array}
\:\:\:\: ,
\label{ele.hop}
\ee
\noindent
the superconducting gap being given by

\be
\Delta_j = \biggl\{ \begin{array}{l} \Delta e^{ i \frac{\chi}{2} }  \: {\rm for} \: j \in  \{ - \Lambda + 2 , \ldots 0 \} \\
\Delta e^{ - i \frac{\chi}{2} }  \: {\rm for} \:j \in\{ \ell , \ldots , \Lambda + \ell- 2 \} \\
0  \:  {\rm for} \:  j \in \{ 1 , \ldots ,  \ell - 1 \}
                       \end{array}
\:\:\:\: ,
\label{su.gap}
\ee
and the potential by:
\be  V_j = \biggl\{ \begin{array}{l} V_C,{\rm for} \:1 \leq j\leq \ell -1)  \\
                0\:   (\hbox{otherwise}  )   
                    \end{array}.
\ee 
\noindent
The construction
of the function ${\cal G} ( E ; \chi )$ is readily achieved by following the same
procedure we used in the continuum case. However, the (lattice) particle and
hole momenta within C and within the leads are now related to the
(negative) energy  $E$, by means of the  lattice dispersion relations, that is

\begin{eqnarray}
- 2 t_S \cos ( \beta_p ) - \mu &=&  \sqrt{E^2 - \Delta^2} \nonumber \\
-2 t_S \cos ( \beta_h ) - \mu &=&  -  \sqrt{E^2 - \Delta^2} 
\;\;\;\; , 
\label{lat.ex0}
\end{eqnarray}
\noindent
and
\begin{eqnarray}
- 2 J \cos ( \alpha_p) + V_C-\mu &=& E \nonumber \\
  2 J \cos ( \alpha_h ) - V_C +\mu
&=&  E 
\:\:\:\: . 
\label{lat.ex1}
\end{eqnarray}
(In \cite{ACZ}, the definition of $\beta_h$ was shifted by $\pi$.) 
${\cal F}$ and ${\cal G}$ can be defined as in Eqs.  (\ref{rafi.1},\ref{m.6a}) and 
can be seen to possess the properties listed in section \ref{general} with the following modifications. 
The branch cuts now run from $\Delta$ to $\tilde E_S=\sqrt{(2t_S+\mu )^2+\Delta^2}$ 
and from $-\Delta$ to $-\tilde E_S$. Furthermore, if $\tilde E_S<2J+V_C-\mu$ there are 
normal bound states (NBS's) in the energy range $\tilde E_C<|E|<2J+V_C-\mu$ as indicated in 
Fig. (\ref{lat1}). 
Looking at the most general case in which both ABS's and NBS's contribute
to $I^{(0)} [ \chi ]$, we obtain $I^{(0)} [ \chi ] = I^{(0)}_{\rm ABS} [ \chi ] +
 I^{(0)}_{\rm SS} [ \chi ] + I^{(0)}_{\rm NBS} [ \chi ]$, with

\begin{eqnarray}
I^{(0)}_{\rm ABS} [ \chi ] &=&  - \frac{2 e}{2 \pi i } \int_{\Gamma_{\rm ABS}} \: d E \:
 \partial_\chi \{ \ln {\cal G} ( E ; \chi )  \} \nonumber \\
I^{(0)}_{\rm SS} [ \chi ] &=&  - \frac{2 e}{2 \pi i } \int_{\Gamma_{\rm SS}} \: d E \:
 \partial_\chi \{ \ln {\cal G} ( E ; \chi )  \} \nonumber \\
I^{(0)}_{\rm NBS} [ \chi ] &=&  - \frac{2 e}{2 \pi i } \int_{\Gamma_{\rm NBS}} \: d E \:
 \partial_\chi \{ \ln {\cal G} ( E ; \chi )  \}
\:\:\:\: ,
\label{root.91a}
\end{eqnarray}
and the path $\Gamma_{\rm ABS}$ defined as in section \ref{general}, 
$\Gamma_{\rm SS}$ being a path surrounding the branch cut lying over
the real axis from $E = - \bar{E}_S$ to $E = - \Delta$, and 
$\Gamma_{\rm NBS}$ being a path surrounding the NBS's. (See Fig.\ref{lat1}{\bf a}
for a sketch of the integration paths.) (clearly, $I_{\rm NBS}^{(0)} [ \chi ] = 0$ 
if there are no NBS's). Considering the path $\Gamma$ made by $\Gamma_{\rm ABS}
\cup \Gamma_{\rm SS} \cup \Gamma_{\rm NBS}$, all run through  clockwise, and by
the outer closed path $\bar{\Gamma}$ (Fig.\ref{lat1}{\bf b)}), made by the arc   $\Sigma$ closed along
the imaginary axis, as ${\cal G} ( E ; \chi )$ has no poles in the region of
the complex plane bounded by $\Gamma$, sending the radius of $\Sigma$ to infinity, we
readily obtain Eq. (\ref{com.4.az}).  When computing the integral, we consider 
again that, for large $\ell$, we may solve Eqs. (\ref{lat.ex1}) for $\alpha_p , \alpha_h$
with $ E = i \omega$, by setting 

\beq
\alpha_p \approx \alpha_F + \frac{ i \omega}{v_F} \;\;\; , \;\;
\alpha_h \approx  \alpha_F - \frac{i \omega}{v_F} \;\;\;\; , 
\label{lat.ex2}
\eneq
\noindent
with $- 2 J \cos ( \alpha_F ) + V_C - \mu  = 0$ and $v_F = 2 J \sin ( \alpha_F )$.
By direct calculation, one finds that the normal reflection amplitudes at the Fermi
level for particle-like states are given by

\beq
\bar{N}_R^p  = \bar{N}_L^p = - \frac{( e^{ - i \beta} - e^{ - i \alpha_F - \lambda} )(
e^{  i \beta^*} - e^{ - i \alpha_F - \lambda} )}{[ e^{ 2 \beta_I} + e^{ - 2 \lambda} - 2 \cos ( \alpha_F ) \cos ( \beta_R ) e^{ \beta_I - \lambda} ] } 
\:\:\:\: , 
\label{lat.exa1}
\eneq
\noindent
with $\beta \equiv \beta_R + i \beta_I = \beta_p ( E=0) = \beta_h^* (E=0)  $, 
$\cos ( \beta ) = - \frac{\mu}{2 t_S } + i \frac{\Delta}{2 t_S }$, and $(t'')^2 = J t_S e^{ - \lambda}$. 
At variance, the Andreev reflection amplitudes at the Fermi level are given by

\beq
\bar{A}_R^p  = \bar{A}_L^h  =    \frac{2 i e^{ \frac{i}{2} \chi } \sin ( \alpha_F ) \sin ( \beta_R ) 
e^{ \beta_I - \lambda} }{[ e^{ 2 \beta_I} + e^{ - 2 \lambda} - 2 \cos ( \alpha_F ) \cos ( \beta_R ) e^{ \beta_I - \lambda} ] } 
\:\:\:\: . 
\label{lat.exa2}
\eneq
\noindent
As a result, we obtain    that again $I^{(0)}  [ \chi ]$ is given by
Eq. (\ref{com.9}), with $u_\pm ( \chi )$ being
the roots of the equation

\begin{eqnarray}
&u^2& + 1 - 2 u \biggl\{ \hbox{Re} \left[ \frac{ ( e^{ -i  \beta} - e^{ - i\alpha_F - \lambda} )^2
( e^{ i \beta^*} - e^{ - i \alpha_F - \lambda} )^2 e^{ 2 i \alpha_F \ell} }{
[ e^{ 2 \beta_I} + e^{ - 2 \lambda} - 2 \cos ( \alpha_F ) \cos ( \beta_R ) e^{ \beta_I - \lambda} ]^2 }
\right] \nonumber \\
 &-& 4 \cos ( \chi ) \frac{ \sin^2 ( \alpha_F ) \sin^2 ( \beta_R ) e^{ 2 \beta_I - 2 \lambda}}{
[ e^{ 2 \beta_I} + e^{ - 2 \lambda} - 2 \cos ( \alpha_F ) \cos ( \beta_R ) e^{ \beta_I - \lambda} ]^2 }
\biggr\} = 0
\nonumber \\ 
\label{lmo.1}
\end{eqnarray}
\noindent
which implies that   $\vartheta ( \chi )$ is now

\begin{eqnarray}
&~&\vartheta ( \chi )  =  \nonumber \\
&~& {\rm arccos} \biggl\{  - 4 \cos ( \chi ) \frac{ \sin^2 ( \alpha_F ) 
\sin^2 ( \beta_R ) e^{ 2 \beta_I - 2 \lambda}}{
[ e^{ 2 \beta_I} + e^{ - 2 \lambda} - 2 \cos ( \alpha_F ) \cos ( \beta_R ) e^{ \beta_I - \lambda} ]^2 }
\nonumber \\
&&+
 \hbox{Re} \left[ \frac{ ( e^{ - i \beta} - e^{ - i\alpha_F - \lambda} )^2
( e^{ i \beta^*} - e^{ - i \alpha_F - \lambda} )^2 e^{ 2 i \alpha_F \ell} }{
[ e^{ 2 \beta_I }+ e^{ - 2 \lambda} - 2 \cos ( \alpha_F ) \cos ( \beta_R ) e^{ \beta_I - \lambda} ]^2 }
\right] \biggr\}
\:\:\:\: . 
\label{lmo.2}
\end{eqnarray}
\noindent
From Eq. (\ref{lmo.2}) we see that again,   as it happens in the continuum model, also in
the lattice model one may tune the system to the perfect Andreev
point, at which $I^{(0)} [ \chi ]$ takes a sawtooth dependence on $\chi$, 
and that, in order to do so, one needs two tuning parameters.
Indeed, as it appears from Eqs. (\ref{lmo.2}),
a sawtooth formula for $I^{(0)} [ \chi ]$ is achieved once
one sets 
$\lambda + \beta_I = 0$ and $\alpha_F  = \pm \beta_R$.
These conditions are the analogs of the conditions
for the continuum model, with now $t''$ and $V_C$ playing the role
of tuning parameters (a different choice for the tuning
parameters was made in \cite{ACZ}, where an additional
normal scattering potential $V \{ \delta_{j,1} + \delta_{j , \ell - 1} \}$ 
was added at the interfaces, but $V_C$ was set to zero). 

Another important observation is that, 
by setting $\mu = V_C =   0$ (and, accordingly, $\beta_R = \alpha_F = 
\pi / 2$), Eq. (\ref{com.9}) yields
 
\beq
 I^{(0)} = 
-  \frac{ e v_F}{2 \pi \ell} 
 \frac{\partial}{\partial \chi} \left\{ {\rm arccos}^2 \left[ \frac{4 J^2 \Delta_B^2 
\cos ( \chi ) + ( J^2 - \Delta_B^2)^2 }{ ( J^2 + \Delta_B^2 )^2} \right]\right\}
\:\:\:\: ,
\label{jabs.10}
\eneq
\noindent
with 
\beq
\Delta_B = (t'')^2 \left[ \frac{\bar{E}_S - \Delta}{2 t_S^2}
\right]
\:\:\:\: ,
\label{jabs.8}
\eneq
\noindent
and, clearly,  $\bar{E}_S = \sqrt{ 4 t_S^2 + \Delta^2}$. 
Eq. (\ref{jabs.10}) is equal to Eq. (3.13) of \cite{ACZ}, despite
the fact that the latter one was derived within an effective two-boundary model
Hamiltonian, obtained by trading the superconducting leads for effective
boundary interactions which, well below the superconducting gap, 
become nearly energy independent. Indeed, the main result   of our 
formalism is that, in principle, when $\ell \gg 1$, it allows for performing calculations
of the dc Josephson current by just focusing on states near 
the Fermi level. This is crucial in motivating the step of trading the actual interface
model for an effective boundary Hamiltonian, which is much simpler to deal
with, especially in the case of an interacting central region, to which 
our approach is likely to apply, as well \cite{GA}. Within the effective
boundary Hamiltonian formulation one readily sees, for instance, that,  
allowing the central region to host an effectively attractive
interaction between electrons should allow for the system to dynamically
self-tune to  the perfect Andreev reflection point, as $\ell$ becomes large \cite{ACZ}.

\noindent

\noindent

\noindent

\section{Finite-temperature generalization}
\label{fin_t}

It is easy to generalize our contour result to compute the dc Josephson
current at finite temperature $T$, $I [ \chi ; T]$. Now \cite{ishi,been.0,furus}
the contour $\Gamma$ gets deformed into a sum of 
circles around the poles of the Fermi function at $\omega_n=2\pi T(n+1/2)$, yielding
\begin{eqnarray}
&~& I [ \chi ; T]  \approx     2 eT \sum_{n=-\infty}^\infty \times \nonumber \\
&~& \left[ \frac{ \partial_\chi \hbox{ Re} [ \bar{A}_R^p \bar{A}_L^h  ] }{  \cosh \left( \frac{  2\omega_n\ell }{
v_F} \right)     - \hbox{ Re} [ \bar{N}_R^p \bar{N}_L^p e^{ 2 i \alpha_F \ell} 
+ \bar{A}_R^p \bar{A}_L^h ]   } \right] 
\:\:\:\: . \label{sumT2}
\end{eqnarray}
\noindent
Of course, this gives our $T=0$ result at $T\ll v_F/\ell$ where we may approximate the sum by an integral. At $T\gg v_F/\ell$ we 
may approximate the sum by the two terms with $\omega_n=\pm \pi T$:
\be 
 I [ \chi ; T]  \approx  8 
eTe^{-2\pi T\ell/v_F}\partial_\chi \hbox{ Re} [ \bar{A}_R^p \bar{A}_L^h ] +O\left( e^{-6\pi T\ell/v_F}\right).
\label{d.dec}
 \ee
This becomes exponentially small when $T\gg v_F/\ell$. 

\begin{figure}
\centering \includegraphics*[width=0.80\linewidth]{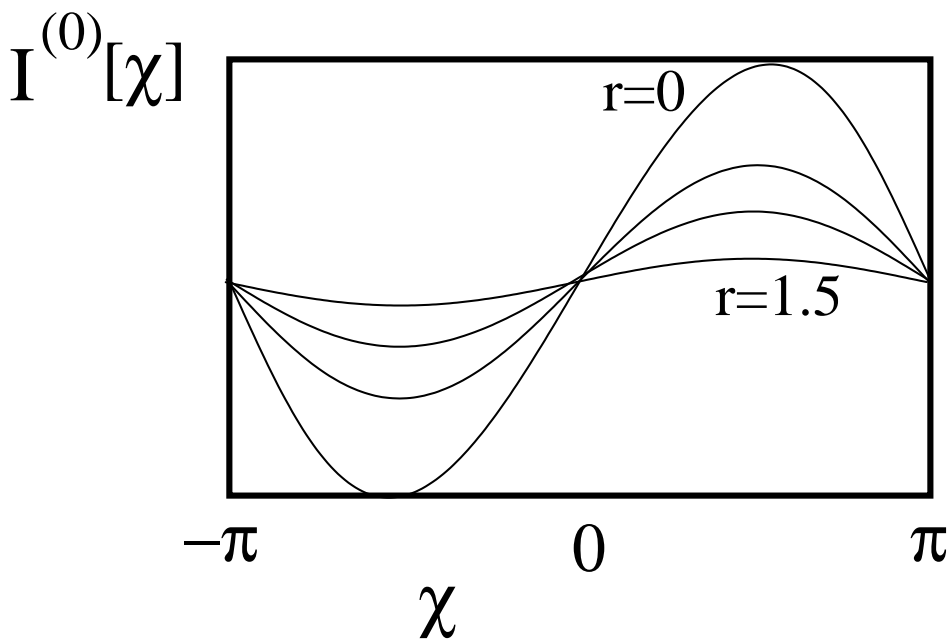}
\caption[]{$I [ \chi ; T]$ versus $\chi$ for the two-interface BTK model for various values of 
$T$, with the system parameters chosen so that $\hbox{Re} \beta_F = 
\hbox{Im} \beta_F  $, $\ell=20$, $v_F = 1$, $Z = .02$, $\alpha_F \ell = 
13  \pi$. The results are displayed for different values of
$r = 4 \pi \ell T / v_F$. From top to bottom, we have
set $r=0,0$ (corresponding to $I^{(0)} [ \chi ]$), 
$r=.35,0.75,1.5$. For $r > 1.5$ the $I [ \chi ; T ]$ 
becomes  negligible, compared to $I^{(0)} [ \chi ]$.}
\label{IT}
\end{figure}

For general values of the ratio  $v_F/T\ell$, the sum in Eq. (\ref{sumT2}) can easily be 
performed numerically. Representative results are shown in Fig. (\ref{IT}), where we report $I [ \chi ; T]$ versus $\chi$
for the BTK  model system we study in section \ref{explicit}
for different values of $T$, at fixed system parameters (see caption). In particular, we see
that there is a rapid reduction of the current as soon as the ratio 
$r = 4 \pi \ell T / v_F \sim 1.5$, which is consistent with
the exponential decay evidenced in Eq. (\ref{d.dec}).

\section{Conclusions}
\label{concl}

By using analytic properties of the scattering matrix for an SNS system, we have expressed the Josephson current, 
which has contributions from both bound and scattering states, as a single contour integral in the complex 
energy plane. In the limit of a long central region, we then proved that the current can be expressed in 
terms of properties of the Andreev bound states at the Fermi energy only: namely the normal and Andreev 
scattering amplitudes. The result holds at finite temperature provided that $T, v_F/\ell \ll \Delta$. This result 
shows that the Josephson current is a {\it universal} quantity, 
in this long length low temperature limit, and justifies the low energy Hamiltonian approach used in \cite{ACZ}, 
which was crucial for treating 
Luttinger liquid interaction effects. It also paves the way towards an extension to the non-equilibrium (AC) 
Josephson effects.

We would like to thank C. W. J. Beenakker, J.-S. Caux, A. Tagliacozzo and A. Zagoskin for helpful discussions and correspondence. 
DG would like to thank the Department of Physics and Astronomy of the
University of British Columbia for the kind hospitality at various stages
of this work. This research was supported in part by NSERC and CIfAR.

\appendix 

\section{Derivation of Eqs.(\ref{amp.12},\ref{ampe.12})}
\label{appe_a}

Throughout our derivation, 
Eqs.(\ref{amp.12},\ref{ampe.12}) of section \ref{long} are quite crucial, 
as they allow us to relate the normal and Andreev reflection amplitudes at the
S-N interfaces to the transmission matrices $L$- and $R$. In this appendix we derive
them in detail, starting from the right-hand S-N interface. By definition, 
the normal and the Andreev reflection amplitudes for a particle-like solution
of the BdG equations at the right-hand interface are defined by considering
a solution that, within C, is given by

\beq
\left[ \begin{array}{c}
        u ( x )  \\ v ( x ) 
       \end{array} \right]_{p,R} =  \left[ \begin{array}{c}
e^{ i \alpha_p (x - \ell) } + N_R^p  ( E )  e^{ - i \alpha_p ( x - \ell) }  \\
A_R^p  ( E ) e^{  i \alpha_h ( x - \ell ) } \end{array}
\right]
\;\;\;\; . 
\label{amp.5}
\eneq
\noindent
Similarly, the normal and the Andreev reflection amplitudes for a hole-like solution
of the BdG equations at the right-hand interface are defined by considering
a solution that, within C, is given by

\beq
\left[ \begin{array}{c}
        u ( x )  \\ v ( x )  
       \end{array} \right]_{h,R} = 
\left[ \begin{array}{c} A_R^h  ( E )  e^{ - i \alpha_p (x - \ell) } \\
  e^{ - i \alpha_h (x - \ell) }   + N_R^h ( E )  e^{  i \alpha_h (x - \ell) } 
       \end{array} \right]
\:\:\:\: . 
\label{amp.6}
\eneq
\noindent
[Note that Eq. (\ref{amp.5}) and (\ref{amp.6}) correspond to a particle or hole respectively, incident on the right interface, 
reflected as a particle or hole.] By definition of the $R$-matrix, we find that, in the
lead R,  the amplitudes corresponding to the solution in Eq. (\ref{amp.5}) 
are  given by
\beq
\left[ \begin{array}{c} A_1^+ \\ A_2^+ \\ A_3^+ \\ A_4^+ 
       \end{array} \right] = 
\left[ \begin{array}{c} R_{1,1}  + R_{1,2} N_R^p  ( E ) + R_{1,4} A_R^p  ( E ) \\
R_{2,1}  + R_{2,2} N_R^p  ( E ) + R_{2,4} A_R^p  ( E ) \\
R_{3,1}  + R_{3,2} N_R^p  ( E )  + R_{3,4} A_R^p  ( E )  \\
R_{4,1}  + R_{4,2} N_R^p  ( E ) + R_{4,4} A_R^p  ( E )   
\end{array} \right]
\:\:\:\: , 
\label{amp.7}
\eneq
\noindent
while the ones corresponding to the solution in Eq. (\ref{amp.6}) are
given by
\beq
\left[ \begin{array}{c} A_1^+ \\ A_2^+ \\ A_3^+ \\ A_4^+ 
       \end{array} \right] = 
\left[ \begin{array}{c} R_{1,2} A_R^h  ( E )    + R_{1,3}    + R_{1,4} N_R^h  ( E )  \\
R_{2,2} A_R^h  ( E )    + R_{2,3}    + R_{2,4} N_R^h  ( E )  \\
R_{3,2} A_R^h  ( E )    + R_{3,3}    + R_{3,4} N_R^h  ( E )  \\
R_{4,2} A_R^h  ( E )   + R_{4,3}    + R_{4,4} N_R^h  ( E )   
\end{array} \right]
\:\:\:\: . 
\label{amp.8}
\eneq
\noindent
As we are eventually interested in computing $I^{(0)} [ \chi ]$ with the
formula in Eq. (\ref{main}), we focus on solutions   with energy $|E| < \Delta$.
Clearly, acceptable solutions in R must not contain terms exponentially growing 
as $x \to \infty$.  According to Eq. (\ref{as.2}),  this implies $A_2^+ = A_4^+ = 0$. The corresponding relations,
derived from Eqs.(\ref{amp.7}) and from Eqs.(\ref{amp.8}), allow for fully
determining $N_R^p ( E ) , A_R^p ( E )$ and $N_R^h ( E ) , A_R^h ( E ) $, 
respectively. As a result, one obtains Eqs.(\ref{amp.12}). 
 
A similar analysis applies to the left-hand S-N interface. By definition, 
the normal and the Andreev reflection amplitude for a particle-like
solution of BdG equations are defined by considering  a   solution that, within C, is given by

\beq
\left[ \begin{array}{c}
        u ( x ) \\ v ( x ) 
       \end{array} \right]_{p,L} =  \left[ \begin{array}{c}
N_L^p  ( E )  e^{ i \alpha_p x} + e^{ - i \alpha_p x}  \\
A_L^p  ( E )  e^{ - i \alpha_h x} \end{array}
\right]
\;\;\;\; . 
\label{amp.1}
\eneq
\noindent
Similarly, the normal and the Andreev reflection amplitude for a hole-like
solution of BdG equations are defined by considering  a   solution that, within C, is given by

\beq
\left[ \begin{array}{c}
        u ( x )  \\ v ( x ) 
       \end{array} \right]_{h,L} = 
\left[ \begin{array}{c} A_L^h  ( E )  e^{ i \alpha_p x } \\
N_L^h  ( E )  e^{ - i \alpha_h x} +         e^{i \alpha_h x}  
       \end{array} \right]
\:\:\:\: . 
\label{amp.2}
\eneq
\noindent
To compute $I^{(0)} [ \chi ]$ one needs the reflection amplitudes at the
Fermi level. Thus, one has to consider solutions of the BdG equations as the
ones in Eqs.(\ref{amp.1},\ref{amp.2}) at   energy
$|E| < \Delta$. Within L, these solutions  must   contain no exponentially growing terms. Thus, we
must set $A_1^- = A_3^- = 0$ in Eq. (\ref{as.1}), getting

\beq
 \left[ \begin{array}{c} N_L^p ( E ) \\ 1 \\ A_L^p ( E ) \\ 0 \end{array} \right]
= \left[ \begin{array}{c} L_{1,2} A_2^- + L_{1,4} A_4^- \\ L_{2,2} A_2^- + L_{2,4} A_4^- \\
   L_{3,2} A_2^- + L_{3,4} A_4^- \\ L_{4,2} A_2^- + L_{4,4} A_4^- \        
         \end{array} \right]
\:\:\:\: , 
\label{ampe.1}
\eneq
\noindent
and

\beq
 \left[ \begin{array}{c} A_L^h ( E ) \\ 0 \\ N_L^h ( E ) \\ 1 \end{array} \right]
= \left[ \begin{array}{c} L_{1,2} A_2^- + L_{1,4} A_4^- \\ L_{2,2} A_2^- + L_{2,4} A_4^- \\
   L_{3,2} A_2^- + L_{3,4} A_4^- \\ L_{4,2} A_2^- + L_{4,4} A_4^- \        
         \end{array} \right]
\:\:\:\: . 
\label{ampe.2}
\eneq
\noindent
Getting rid of $A_2^- , A_4^-$, from Eqs.(\ref{ampe.1},\ref{ampe.2}) one eventually
obtains  Eqs.(\ref{ampe.12}) for $N_L^p ( E )$ and  $A_L^p ( E )$.

\section{Derivation of the normal and Andreev reflection amplitudes in the examples
of section \ref{explicit}}
\label{na_expl}

We now outline the calculation of the reflection amplitudes used to compute $I^{(0)} [ \chi ]$ for
the  two-interface continuum model  and for the lattice model   studied in
section \ref{explicit}. 

The calculation of the reflection amplitudes in the continuum model can be performed
within the framework of BdG equations, as discussed in \cite{btk}. To be specific, let
us focus on the right-hand S-N interface. For the model we consider in section \ref{explicit},
this is described by assuming a coordinate-dependent
gap function $\tilde{\Delta} ( x ) $ and a potential $\tilde{V} ( x )$ given by

\begin{eqnarray} 
 \tilde{\Delta} ( x ) &=& \Delta e^{ - \frac{i}{2} \chi} \theta ( x - \ell ) \nonumber \\
\tilde{V} ( x ) &=& V_0 \delta ( x - \ell )  + V_C \theta ( \ell - x )
\:\:\:\: . 
\label{appeb.1}
\end{eqnarray}
\noindent
The wavefunctions $u ( x ) , v ( x )$ for a state with energy $E$ must solve 
the time-independent BdG equations

\begin{eqnarray}
E u ( x ) &=&  \left[ - \frac{1}{2m} \frac{d^2}{d x^2} - \mu + \tilde{V} ( x ) \right] u ( x ) + \tilde{\Delta} ( x ) v ( x )  \nonumber \\
 E v ( x ) &=& \tilde{\Delta}^* ( x ) u ( x ) -   \left[ - \frac{1}{2m} \frac{d^2}{d x^2} - \mu + \tilde{V} ( x ) \right] v ( x ) 
\:\:\:\: . 
\label{appeb.2}
\end{eqnarray}
\noindent
To compute $N_R^p ( E ) , A_R^p ( E )$ we require $ u ( x ) , v ( x )$ to obey
boundary conditions such that, within C, they are given by Eq. (\ref{amp.5}), with $\alpha_p = 
\sqrt{ 2 m ( E + \mu - V_C )}$ and $\alpha_h = \sqrt{ 2 m ( - E + \mu - V_C )}$, while, within
R, they are given by Eq. (\ref{as.2}), with $B_2 = B_4 = 0$. In order to
obey Eqs. (\ref{appeb.2}), at the interface $u ( x ) , v ( x ) $ must satisfy the
continuity condition \cite{btk}

\beq
\left[ \begin{array}{c} u ( \ell^- ) \\ v ( \ell^- ) \end{array} \right] = 
\left[ \begin{array}{c} u ( \ell^+ ) \\ v ( \ell^+ ) \end{array} \right]
\:\:\:\: , 
\label{appeb.3}
\eneq
\noindent
and, in addition, the discontinuity in the derivatives must cancel the term
due to the localized normal scattering potential at the interface, that is

\beq
\left[ \begin{array}{c} u^{'} ( \ell^- ) \\ v^{'} ( \ell^- ) \end{array} \right] - 
\left[ \begin{array}{c} u^{'} ( \ell^+ ) \\ v^{'} ( \ell^+ ) \end{array} \right] 
= - Z \left[ \begin{array}{c} u ( \ell  ) \\ v ( \ell  ) \end{array} \right]
\:\:\:\: , 
\label{appeb.4}
\eneq
\noindent
with $Z = 2 m V_0$. Eqs.(\ref{appeb.3},\ref{appeb.4}) provide a set of four algebraic
equations in the four unknowns $B_1 , B_3 , N_R^p ( E ) , A_R^p ( E )$. Getting rid
of $B_1 , B_3$, one obtains a set of two algebraic equations in the unknowns 
$N_R^p ( E ) , A_R^p ( E)$. Solving the resulting equations for $E= 0 $, one obtains 
the formulas we give in Eqs.(\ref{r.test.1},\ref{r.test.2}). The amplitudes 
$N_R^h ( E ) , A_R^h (E)$ are obtained following the same procedure, but using 
the formula in Eq. (\ref{amp.6}) for the solution within C. In the same way, looking for
solutions of Eqs. (\ref{appeb.2}), but now with $\tilde{\Delta} ( x ) = 
\Delta e^{ \frac{i}{2} \chi} \theta ( - x )$ and 
$\tilde{V} ( x ) = V_0 \delta ( x ) + V_C \theta ( x )$, one obtains
the particle-like and the hole-like reflection amplitudes at the left-hand interface. 

To derive the normal and Andreev reflection amplitudes in the lattice model, one
may follows exactly the same procedure discussed above for the continuum model, except
that, now, the lattice BdG equations in Eq. (\ref{csns.1}) must be used. For instance, in
order to compute $N_R^p ( E ) , A_R^p ( E ) $ in the lattice model, one considers
 Eqs.(\ref{csns.1}) with the hopping given by

\be
\tau_{j , j+1} = \biggl\{ \begin{array}{l} J \: {\rm for} j \leq  \ell - 2  \\
t_S \: {\rm for}  \:j \geq \ell \\
t'' \: {\rm for} \:  j =  \ell - 1
                       \end{array}
\:\:\:\: ,
\label{ele.hop.b}
\ee
\noindent
the superconducting gap given by

\be
\Delta_j = \biggl\{ \begin{array}{l}0  \:  {\rm for} \:  j \leq   \ell - 1   \\
\Delta e^{ - i \frac{\chi}{2} }  \: {\rm for} \:j \geq \ell  \\

                       \end{array}
\:\:\:\: ,
\label{su.gap.b}
\ee
and the potential by:
\be  V_j = \biggl\{ \begin{array}{l} V_C,{\rm for}   j\leq \ell -1)  \\
                0\:   (\hbox{otherwise}  )   
                    \end{array}.
\label{po.b}
\ee 
\noindent
For $j \leq \ell - 1$, the solution with energy $E$ obeying the appropriate boundary condition is given by

\beq
\left[ \begin{array}{c} u_j \\ v_j \end{array} \right]_{p,R} = 
\left[ \begin{array}{c} e^{ i \alpha_p ( j - \ell ) } + N_R^p e^{ - i \alpha_p ( j - \ell ) } \\
        A_R^p e^{ - i \alpha_h ( j - \ell ) }
       \end{array} \right]
\:\:\:\: , 
\label{appec.1}
\eneq
\noindent
with $\alpha_p , \alpha_h$ given in Eq. (\ref{lat.ex1}). Within R, one rather obtains

\beq
\left[ \begin{array}{c} u_j \\ v_j \end{array} \right]_{S} = 
B_1 \left[ \begin{array}{c} \cos \left( \frac{\Psi}{2} \right) \\ 
           -  e^{ \frac{i}{2} \chi}  \sin \left( \frac{\Psi}{2} \right)
           \end{array} \right] e^{ i \beta_p j } + 
B_3 \left[ \begin{array}{c}-  e^{ - \frac{i}{2} \chi}  \sin \left( \frac{\Psi}{2} \right) \\
 \cos \left( \frac{\Psi}{2} \right) 
           \end{array} \right] e^{ - i \beta_h j } 
\;\;\;\; , 
\label{appec.2}
\eneq
\noindent
with $\beta_p , \beta_h$ given in Eq.(\ref{lat.ex0}) and 

\beq
 \cos ( \Psi ) = \frac{ 2 t_S \cos ( \beta_p ) + \mu}{E}  = \frac{ - 2 t_S \cos ( \beta_h ) - \mu}{E} \;\;\; , 
\;\;
\sin ( \Psi ) = \frac{\Delta}{E}
\:\:\:\: . 
\label{appec.3}
\eneq
\noindent
As discussed in detail in \cite{ACZ}, in the lattice interface model described by 
Eqs.(\ref{csns.1},\ref{ele.hop.b},\ref{su.gap.b},\ref{po.b}), the matching conditions at
the interface, given in Eqs.(\ref{appeb.3},\ref{appeb.4}) for the continuum model, are
just substituted by the lattice equations for $u_{\ell - 1} , v_{\ell - 1}$ and for $u_\ell , v_\ell$.
Requiring the solution in Eqs.(\ref{appec.1},\ref{appec.2}) to satisfy Eqs.(\ref{csns.1})
for $u_{\ell - 1} , v_{\ell - 1}$ and for $u_\ell , v_\ell$ and getting rid of $B_1 , B_3$, one
obtains a set of equations for $N_R^p , A_R^p$, whose solutions, at the Fermi level, provide 
us with the result in Eqs.(\ref{lat.exa1},\ref{lat.exa2}). Following the same procedure, after
substituting $\left[ \begin{array}{c} u_j \\ v_j \end{array} \right]_{p,R}$ in Eq. (\ref{appec.1}) with

\beq
\left[ \begin{array}{c} u_j \\ v_j \end{array} \right]_{h,R} = 
\left[ \begin{array}{c}  A_R^h e^{ - i \alpha_p ( j - \ell ) } \\
e^{ - i \alpha_h ( j - \ell ) } + N_R^h e^{  i \alpha_h ( j - \ell ) } \\
              \end{array} \right]
\:\:\:\: , 
\label{appec.z}
\eneq
\noindent
one obtains $N_R^h , A_R^h$. Similarly, one readily obtains $N_L^p , A_L^p$ and
$N_L^h , A_L^h$, as well. 

\section{Derivation of Eq.(\ref{mma.1}) and of Eq.(\ref{mudef})}
\label{derivat}

In this appendix, we outline the derivation of Eq.(\ref{mma.1}), which is
crucial for our proof. To do so, we have to go through the derivation of 
Eq.(\ref{mudef}), which gives the necessary condition to be satisfied
by solutions of BDG equations in a one-dimensional box   defined by 
$- \frac{L}{2} < x < \ell + \frac{L}{2}$.  Thus, 
we will also outline the procedure for deriving the  formulas
for the coefficients $\mu_{n,m}$. The starting point is noticing that a
solution of the BDG equations in the one-dimensional box,  $\left[ \begin{array}{c}u( x ) \\ v ( x )
                                                                \end{array} \right]$,
must obey vanishing boundary conditions at both boundaries of the box, that
is $\left[ \begin{array}{c} u \left( - \frac{L}{2} \right) \\
            v \left( - \frac{L}{2} \right)            
                      \end{array} \right] = \left[ \begin{array}{c} u \left(\frac{L}{2} \right) \\
            v \left(  \frac{L}{2}\right)            
                      \end{array} \right] = 0$. Clearly, $\left[ \begin{array}{c}u( x ) \\ v ( x )
                                                                \end{array} \right]$
takes the form  given in Eq. (\ref{as.1}) for $x \to - \frac{L}{2}$ and the
form given in  Eq. (\ref{as.2}) for 
$x \to \ell +  \frac{L}{2} $. Thus, vanishing boundary conditions imply the system
of algebraic equations

\begin{eqnarray}
  \cos ( \Psi / 2 ) [ A_1^- e^{ - \frac{i}{2} \beta_p L} +  A_2^- e^{   \frac{i}{2} \beta_p L} ]
- e^{ i \chi / 2 } \sin ( \Psi / 2 )  [ A_3^- e^{  \frac{i}{2} \beta_h L} +  A_4^- e^{  - \frac{i}{2} \beta_p L} ]
 &=&  0 \nonumber \\
 - e^{ - i \chi / 2 } \sin ( \Psi / 2 )   [ A_1^- e^{ - \frac{i}{2} \beta_p L} +  A_2^- e^{   \frac{i}{2} \beta_p L} ]
+ \cos ( \Psi / 2 ) [ A_3^- e^{  \frac{i}{2} \beta_h L} +  A_4^- e^{  - \frac{i}{2} \beta_p L} ]
 &=&  0 \; ,  \nonumber \\
\label{ap.1}
\end{eqnarray}
\noindent
and

\begin{eqnarray}
 \cos ( \Psi / 2 ) [ A_1^+ e^{   \frac{i}{2} \beta_p L} +  A_2^+ e^{  - \frac{i}{2} \beta_h L} ]
- e^{ - i \chi / 2 } \sin ( \Psi / 2 )  [ A_3^+ e^{  -\frac{i}{2} \beta_h L} +  A_4^+ e^{    \frac{i}{2} \beta_h L} ]
&=& 0 \nonumber \\
 - e^{   i \chi / 2 } \sin ( \Psi / 2 )   [ A_1^+ e^{   \frac{i}{2} \beta_p L} +  A_2^+ e^{ -  \frac{i}{2} \beta_h L} ]
+ \cos ( \Psi / 2 ) [ A_3^+ e^{  -\frac{i}{2} \beta_h L} +  A_4^+ e^{    \frac{i}{2} \beta_h L} ]
&=& 0 \: . \nonumber \\
&&  
\label{ap.2}
\end{eqnarray}
\noindent
By definition of the transmission matrix $M$, one finds that
$A_1^+ , A_2^+ , A_3^+ , A_4^+$ are related to 
$A_1^- , A_2^- , A_3^- , A_4^-$ by  
$\vec{A}^+ = M \vec{A}^-$. Using this last equation to 
express $A_1^+ , A_2^+ , A_3^+ , A_4^+$ in terms  of $A_1^- , A_2^- , A_3^- , A_4^-$,
the system given by Eqs.(\ref{ap.1},\ref{ap.2})  can be traded for a homogeneous $4 \times 4$ system
in the unknowns $A_1^- , A_2^- , A_3^- , A_4^-$, given by the equations

\begin{eqnarray}
 e^{ - \frac{i}{2} \beta_p L} A_1^- + e^{ \frac{i}{2} \beta_p L} A_2^- &=& 0 \nonumber \\
e^{  \frac{i}{2} \beta_h L} A_3^- + e^{ - \frac{i}{2} \beta_h L} A_4^- &=& 0
\:\:\:\: ,
\label{ap.3}
\end{eqnarray}
\noindent
and

\begin{eqnarray}
 \{ e^{ \frac{i}{2} \beta_p L}  M_{1,1} +  e^{ - \frac{i}{2} \beta_p L} M_{2,1} \} A_1^- 
+  \{ e^{ \frac{i}{2} \beta_p L}  M_{1,2} +  e^{ - \frac{i}{2} \beta_p L }M_{2,2} \} A_2^- &+& \nonumber \\
  \{ e^{ \frac{i}{2} \beta_p L}  M_{1,3} +  e^{ - \frac{i}{2} \beta_p L }M_{2,3} \} A_3^- 
 +  \{ e^{ \frac{i}{2} \beta_p L}  M_{1,4} +  e^{ - \frac{i}{2} \beta_p L} M_{2,4} \} A_4^- 
&=& 0 \nonumber \\
 \{ e^{ - \frac{i}{2} \beta_h L}  M_{3,1} +  e^{   \frac{i}{2} \beta_h L} M_{4,1} \} A_1^- 
+  \{ e^{ -\frac{i}{2} \beta_h L}  M_{3,2} +  e^{   \frac{i}{2} \beta_h  L} M_{4,2} \} A_2^- &+& \nonumber \\
   \{ e^{ - \frac{i}{2} \beta_h L}  M_{3,3} +  e^{   \frac{i}{2} \beta_h  L} M_{4,3} \} A_3^- 
 +  \{ e^{-  \frac{i}{2} \beta_h L}  M_{4,4} +  e^{   \frac{i}{2} \beta_h L} M_{4,4} \} A_4^- 
&=& 0  
\:\:\:\: . 
\label{ap.4}
\end{eqnarray}
\noindent
Eqs.(\ref{ap.3},\ref{ap.4}) constitute a system of the form ${\cal M} \vec{A}^- = 0$, 
with the matrix ${\cal M}$ being a known function of the energy $E$ and of 
the transmission matrix elements. Nonzero solutions for $\vec{A}^-$ are then 
only obtained if ${\rm det} [ {\cal M} ] =0$. This latter condition 
yields Eq.(\ref{mudef}) and explicitly defines the coefficients $\mu_{n,m}$.
In particular, by direct calculation one readily obtains

\beq
\frac{\mu_{1,-1}}{\mu_{-1,1}} = \frac{M_{1,1} M_{3,3} - M_{1,3} M_{3,1}}{M_{2,2} M_{4,4} -
M_{2,4} M_{4,2}} 
\:\:\:\: . 
\label{ap.5}
\eneq
\noindent
In order to explicitly show that ${\rm det} [ S ]$ is equal to the right-hand side
of Eq.(\ref{ap.5}), we now perform a different manipulation on the system of
Eqs.(\ref{ap.1},\ref{ap.2}). In particular, we use the definition of the
$S$-matrix, 

\beq
\left[ \begin{array}{c} \sqrt{v_p} A_1^+ \\ \sqrt{v_p} A_2^- \\ \sqrt{v_h} A_3^+ \\ \sqrt{v_h} A_4^-
       \end{array} \right] = S \left[ \begin{array}{c} \sqrt{v_p} A_1^- \\ \sqrt{v_p} A_2^+  \\ \sqrt{v_h} A_3^- \\ \sqrt{v_h} 
A_4^+
       \end{array} \right]
 \;\;\;\; , 
\label{ap.6}
\eneq
\noindent
with $v_{p/h} = \left| \frac{d E }{ d \beta_{p/h}} \right|$, to express 
$A_1^+ , A_2^- , A_3^+ , A_4^-$ in terms of $A_1^- , A_2^+ , A_3^-, A_4^+$.
As a result, one obtains an algebraic homogeneous system of the form

\beq
{\cal M}^{'} \left[ \begin{array}{c}  A_1^- \\  A_2^+  \\   A_3^- \\   A_4^+
       \end{array} \right] = 0 
\:\:\:\: . 
\label{ap.7}
\eneq
\noindent
Eq.(\ref{ap.7}) takes nonzero solutions only if 

\beq
{\rm det} [ {\cal M}^{'} ] = 0 =  \sum_{n , m = \pm 1} \nu_{n,m} e^{ i [ n \beta_p + m \beta_h ] L } 
+ \nu_{0,0}
\:\:\:\: . 
\label{ap.8}
\eneq
\noindent
Comparing Eq.(\ref{ap.8}) with Eq.(\ref{mudef}), one sees that the $\nu_{n,m}$'s must
be equal to the $\mu_{n,m}$'s, apart for an over-all multiplicative constant. 
In particular, this implies

\beq
\frac{\mu_{1,-1}}{\mu_{-1,1}} = \frac{\nu_{1,-1}}{\nu_{-1,1}} \Rightarrow 
{\rm det} [ S ] = \frac{M_{1,1} M_{3,3} - M_{1,3} M_{3,1}}{M_{2,2} M_{4,4} -
M_{2,4} M_{4,2}} 
\;\;\;\;  ,
\label{ap.9}
\eneq
\noindent
that is, Eq.(\ref{mma.1}).

\end{document}